\documentclass[aps,preprint,superscriptaddress,nofootinbib,preprintnumbers]{revtex4-1} 
\usepackage{amsmath,amssymb,amsfonts}
\usepackage{bm, slashed}
\usepackage{color,graphicx}

\usepackage{xcolor}

\definecolor{hgreen}{rgb}{0,0.3,0}
\definecolor{hred}{rgb}{0.3,0,0}
\definecolor{hblue}{rgb}{0,0,0.3}
\usepackage[colorlinks=true,
            linkcolor=hblue,
            citecolor=hgreen,
            filecolor=hblue,
            urlcolor=hred]{hyperref}

\newcommand{\ctilde}{\tilde{c}}
\newcommand{\hgg}{h \to \gamma\gamma}
\newcommand{\qqgg}{q\bar{q} \to \gamma\gamma}
\newcommand{\beq}{\begin{equation}}
\newcommand{\eeq}{\end{equation}}
\newcommand{\bea}{\begin{align}}
\newcommand{\eea}{\end{align}}
\newcommand{\bS}{\mathcal{S}}

\newcommand{\Pp}[1]{p_{#1_+}}
\newcommand{\Pm}[1]{p_{#1_-}}

\newcommand{\agbr}[2]{\langle #1~#2 \rangle}
\newcommand{\sqbr}[2]{[ #1 ~ #2]}
\newcommand{\sigbr}[4]{ \langle #2^#1| #3 | #4^#1 \rangle}
\newcommand{\rs}{{rs}}

\newcommand{\Em}{\mathcal{M}}
\newcommand{\mcA}{\mathcal{A}}
\newcommand{\BH}{\text{BH}}
\newcommand{\dPS}{d{\textrm{\sc ps}}}

\newcommand{\mcG}{\mathcal{G}}

\newcommand{\BAv}[1]{\langle \mathcal{B} \rangle_{#1} / \langle \mathcal{A} \rangle_{#1}}

\def\Cincy{Department of Physics, University of Cincinnati,\\ Cincinnati, OH 45221,USA}
\def\Fermilab{Theoretical Physics Department, Fermilab,\\ P.O. Box 500, Batavia, IL 60510}
\def\Cornell{Department of Physics, LEPP, Cornell University,\\ Ithaca, NY 14853, USA}
\def\Beijing{State Key Laboratory of Theoretical Physics and Kavli Institute for Theoretical Physics China (KITPC), Institute of Theoretical Physics, Chinese Academy of Sciences,\\ Beijing 100190, P. R. China}
\def\UCBerkeley{Department of Physics, University of California,\\ Berkeley, CA 94720, USA}
\def\LBNL{Ernest Orlando Lawrence Berkeley National Laboratory,\\ University of California, Berkeley, CA 94720, USA}

\allowdisplaybreaks[1]

\begin{document}

\title{Probing CP Violation in \texorpdfstring{$h\to \gamma\gamma$}{hgg} with Converted Photons}

\author{Fady Bishara}
\email[Electronic address:]{bisharfy@mail.uc.edu}
\affiliation{\Cincy}
\affiliation{\Fermilab}
\author{Yuval Grossman}
\email[Electronic address:]{yg73@cornell.edu} 
\affiliation{\Cornell}
\author{Roni Harnik}            
 \email[Electronic address:]{roni@fnal.gov}
\affiliation{\Fermilab}
\author{Dean J. Robinson}
\email[Electronic address:]{djrobinson@berkeley.edu} 
\affiliation{\UCBerkeley}
\affiliation{\LBNL}

\author{Jing Shu}
\email[Electronic address:]{shujingtom@gmail.com} 
\affiliation{\Beijing}
\author{Jure Zupan} 
\email[Electronic address:]{zupanje@ucmail.uc.edu} 
\affiliation{\Cincy}

\date{\today}

\begin{abstract}
We study Higgs diphoton decays, in which both photons undergo nuclear conversion to electron-positron pairs. The kinematic distribution of the two electron-positron pairs may be used to probe the CP violating (CPV) coupling of the Higgs to photons, that may be produced by new physics. Detecting CPV in this manner requires interference between the spin-polarized helicity amplitudes for both conversions. We derive leading order, analytic forms for these amplitudes. In turn, we obtain compact, leading-order expressions for the full process rate. While performing experiments involving photon conversions may be challenging, we use the results of our analysis to construct experimental cuts on certain observables that may enhance sensitivity to CPV. We show that there exist regions of phase space on which sensitivity to CPV is of order unity. The statistical sensitivity of these cuts are verified numerically, using dedicated Monte-Carlo simulations.
\end{abstract}

\preprint{FERMILAB-PUB-13-540-T}

\maketitle
\tableofcontents

\section{Introduction}

The recent discovery of a Higgs-like boson \cite{Chatrchyan:2012ufa,Aad:2012tfa} has  prompted intense interest in the precise measurement of its couplings and properties. Such measurements are a direct probe of new physics (NP) beyond the Standard Model (SM), especially since many extensions of the SM modify the Higgs couplings to gauge bosons and fermions. 

Of particular interest is a search for parity and CP violating Higgs decays, since these would be a clear signal of NP \cite{Voloshin:2012tv,Chen:2013ejz,Anderson:2013afp,Sun:2013yra,Boughezal:2012tz,Bolognesi:2012mm,Harnik:2013aja,Berge:2013jra,DellAquila:1988fe,DellAquila:1988rx,Grzadkowski:1995rx,Bower:2002zx,Worek:2003zp,Desch:2003mw,Berge:2008dr,Berge:2008wi,Berge:2011ij,Brod:2013cka}. In Higgs decays to vector bosons the CP violating effects can only be due to irrelevant NP operators. Fortunately, in the SM $\hgg$ (and $h\to Z\gamma$) decays are also due to irrelevant operators, with the first non-zero contribution occurring at one-loop. Thus we can expect large CP violating effects from weak scale NP in $\hgg$. In contrast, the $h \to ZZ^*$ and $h\to WW^*$ decays proceed in the SM through relevant tree-level operators tightly related to the $Z$ and $W$ masses. CP violating effects from NP are expected to be comparatively small in these decay modes.

In the presence of CPV, the total $\hgg$ decay rate must be proportional to the sum of squares of CP-even and CP-odd terms -- i.e $|{\rm CP}_{\rm even}|^2 + |{\rm CP}_{\rm odd}|^2 $ -- and therefore, by comparing the $\hgg$ rate to the SM expectation, one may probe for NP directly.  However, this type of search cannot distinguish between CP-even and CP-odd NP contributions to the total rate. Moreover if the CP-odd contribution is small, then CPV signals are quadratically suppressed, and if it so happens that NP enters both the CP even and odd terms, such that the total $\hgg$ rate matches the SM expectation, then one cannot detect NP at all.  Probing the differential $\hgg$ rate for CPV ameliorates these problems. In the first instance, the differential rate may feature an interference term of the form $2 {\rm CP}_{\rm even}{\rm CP}_{\rm odd}$. Combined with non-interfering terms, one may distinguish CP-even and CP-odd NP contributions. Secondly, small CPV signals are only linearly suppressed in this interference term.

\begin{figure}\centering
\includegraphics[height=4.8cm]{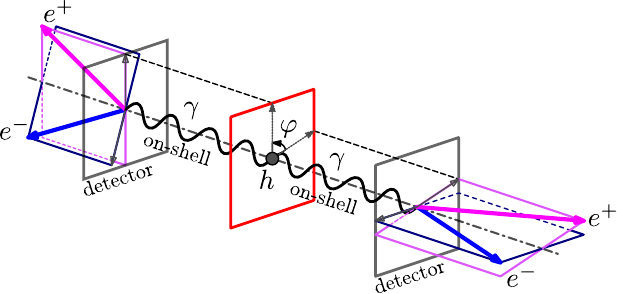}
\caption{An illustration of an example of a CPV sensitive observable in $h\to \gamma\gamma \to 4e$. The Higgs decays to on-shell photons which convert in the detector. The distribution of the azimuthal angle~$\varphi$  between the two planes formed by each positron and its parent photon depends on the Higgs couplings to CP even and odd operators.  The electrons do not need to be co-planar with the corresponding photon-positron planes. The positron-photon plane is shown in magenta and the electron-photon plane in blue. For further details and subtleties see the main text.}
\label{fig:cartoon}
\end{figure}

The $\hgg$ phase space distribution alone, however, is not sensitive to CP violating effects, since the Higgs decays isotropically to two photons. Nevertheless, the underlying CPV structure in the differential $\hgg$ rate may be determined if one is able to measure the linear polarizations of the outgoing photons. This in itself is an old idea, first proposed for the determination of the $\pi^0$ parity~\cite{Yang:1950sr,Yang:1950pe,Kroll:1955ip} and, more recently, to probe NP effects in radiative $B$ decays~\cite{Grossman:2000rk}. It relies on the fact that a spin-$0$ particle decays to either two positive or two negative helicity photon states, which acquire a relative CPV phase in the presence of non-trivial CP structure. The linearly polarized photon states are a superposition of both helicities, permitting one to extract this CPV phase.  It is not feasible to directly measure the linear polarization of ${\mathcal O}(60$~GeV) photons from Higgs decay. However, in both the ATLAS \cite{ATLAS:2012ad} and CMS  \cite{Chatrchyan:2012tw} detectors roughly half of the photons from Higgs decays convert via the well-known Bethe-Heitler (BH) process into $e^+ e^-$ pairs inside the silicon tracker. This has an important benefit: the orientation of the produced $e^+e^-$ pairs encodes the underlying CP structure. Figure \ref{fig:cartoon} illustrates an observable expected to be sensitive to CPV.

Previous proposals to measure CPV in $\hgg$, or in other neutral meson diphoton decays, via double photon conversion appear in Refs. \cite{Kroll:1955ip,Cottingham:1995sh,Voloshin:2012tv}. We extend these studies by performing an analysis of the actual manner in which the leptonic phase space encodes the CP violating effects. We examine the encoding of CPV in the doubly converted Higgs diphoton decay -- hereafter the Higgs-Bethe-Heitler (HBH) process -- both analytically and numerically, making use of the spinor-helicity formalism to obtain compact expressions for the full differential scattering rate and its leading order terms. 

A key difficulty in extracting CPV signals from differential scattering rates like $d\Gamma/d\varphi$, defined in Fig. \ref{fig:cartoon}, is that the signal is largely washed out under integration over the other phase space variables. However, in this work we use our analytic control of the full differential scattering rate to show that large CPV signals may be achieved. We find certain observables exhibit $\mathcal{O}(1)$ sensitivity to CP violating effects on small regions of the phase space, corresponding to a small fraction of the converted decays. These regions are identified by sensitivity parameters, derived from our analysis. We show analytically and confirm numerically that if one cuts the HBH event data according to these sensitivity parameters, the CPV signal is dramatically improved.

Performing such an experiment will be difficult. For example, one must resolve the opening angles in the photon conversion, and this requires extremely accurate tracking resolution. Other subtle effects, such as soft scatterings of the electron and positron in the detector material need also to be examined. In addition, reaching a sensitivity to Higgs couplings would require obtaining a signal-rich sample of $h\to\gamma\gamma$ events. Our approach here is to defer these considerations to future work, and consider mainly the theoretical aspects of this process. In doing this, our intent is to motivate a very challenging measurement, perhaps to be done after the LHC upgrade or in a future Higgs factory. We note however, that as a warm-up to Higgs studies, photon conversions can be studied in background samples as a test of the standard model (we present the phase space structure of $\bar q q \to \gamma\gamma$ with converted photons in an appendix).

This paper is structured as follows. In Section \ref{sec:HDDCPV} we first present a motivating phenomenological analysis of the expected size of CP violation in Higgs decays, followed by an analysis of the unconverted $\hgg$ process. This is presented in the language of helicity amplitudes and we show that the CPV terms in the differential rate arise from helicity amplitude interference. In Section \ref{sec:BHPC} we discuss photon propagation and conversion in modern detectors, angular resolution limits, and the central role of the nuclear form factor in Bethe-Heitler conversion.  In Section \ref{sec:HBHP} we then proceed to examine the HBH process itself, presenting explicit results within the spinor-helicity formalism for the leading order HBH square amplitude. This is followed in Section \ref{sec:O} by a derivation of CPV observables and their sensitivity parameters for the special case that the Higgs is at rest. These sensitivity parameters can be used as cuts, which extract the phase space regions on which we expect $\mathcal{O}(1)$ CPV effects. Numerical simulations are presented which confirm these expectations, and further compare the performance of the various sensitivity parameter cuts. In this work we focus on the $\hgg$ decay, but our analysis can be used to examine other searches, e.g., for $h\to Z\gamma$ or other decays involving converted photons. 

\section{Higgs diphoton decay  with CP violation}
\label{sec:HDDCPV}
\subsection{Motivation for measuring CP-violation in di-photons}
\label{sec:CPV}
The CP structure of Higgs decays was already studied experimentally in $h\to ZZ^*$ decays \cite{ATLAS-CONF-2013-013,Chatrchyan:2012jja}, with pure pseudo-scalar coupling disfavored at more than $3\sigma$ \cite{Chatrchyan:2012jja}. Still, there is strong motivation to measure CPV also in loop induced $\hgg$ decays.  The motivation is based on the expected sizes of CP violating and CP conserving terms in $\hgg$ and $h \to ZZ^*$ decays.  After electroweak symmetry breaking (EWSB), the relevant terms in the effective Lagrangian at the scale $\mu \simeq m_h/2$ are
\begin{align}
	\mathcal{L}_{\rm eff}
	& \supset 
	c_V \frac{m_Z^2}{v} h Z^\mu Z_\mu+c  \frac{\alpha}{\pi v} h F_{\mu\nu}F^{\mu\nu}+c_{Z Z} \frac{\alpha}{\pi v} h Z^{\mu\nu}  Z_{\mu\nu}\notag\\
	& \quad +\tilde c_{Z Z} \frac{\alpha}{2 \pi v} h Z_{\mu \nu} \tilde Z^{\mu\nu} + \ctilde \frac{\alpha}{2 \pi v} hF_{\mu\nu}\tilde F^{\mu\nu}, \label{eqn:LEFF}
\end{align}
in which $F_{\mu\nu}$ and $Z_{\mu\nu}$ are respectively the photon and $Z$ field strengths, and $\tilde{X}^{\mu\nu} = \epsilon^{\mu\nu\alpha\beta}X_{\alpha\beta}$, $\epsilon^{0123} = 1$, while $v = 246$ GeV is the Higgs vev.  Taking Higgs to be a scalar, the first line of  Eq.~\eqref{eqn:LEFF} contains CP even and the second line CP odd operators.\footnote{Since all the couplings are C even, P and CP violation are the same. While we hereafter always refer to CP violation, it should be kept in mind that it is equivalent to parity violation for this effective theory.} Present data imply $c_V =1.04\pm0.13$ \cite{Belanger:2013xza}, assuming CP conservation.  If either $\tilde c_{ZZ}$ or $\ctilde$ are found to be non-zero,\footnote{In fact, $\ctilde$ may be non-zero at three loops in the SM \cite{Czarnecki:1996rx}. However, since this contribution falls well below the feasible detection threshold for the experiments under consideration in this work, we neglect this contribution, and treat the SM value as $\ctilde = 0$.} CPV in Higgs couplings and thus NP will be discovered. 
The couplings $c$ and  $\ctilde $ arise at one-loop in perturbative UV theories, and can be at most ${\mathcal O}(1)$ in order to agree with the observed $h\to \gamma\gamma$ rate. For example, $\ctilde$ can arise from a massive NP fermion loop that is axially coupled to the Higgs. Note that in Eq.~\eqref{eqn:LEFF} we integrated out both $W$ and $t$ loop contributions to $c $, so that $c_{\rm SM}=-0.81$ in the SM. Similarly, the dimension 5 couplings of Higgs to $ZZ$, $c_{ZZ}$ and $\tilde c_{ZZ}$, also arise at one-loop order.  In generic NP models we thus expect $c \sim c_{ZZ}$ and $\tilde c\sim \tilde c_{ZZ}$.

The $h\to ZZ^*$ decay is dominated by the CP even renormalizable coupling $c_V$, while the $h\to \gamma\gamma$ decay is given by higher dimensional operators. The relative size of CP violating effects in any channel is given by the ratio of the interference terms to the total amplitude squared. For $h\to ZZ^*$ the CP odd interference term is proportional to $(\alpha/2\pi) \tilde c_{ZZ} c_V$, while the total squared amplitude is dominated by $c_V^2$. The typical size of CPV observables in $h\to ZZ^*$ is therefore set by the ratio of the two,
\begin{equation}
r_{ZZ^*}= \frac{\alpha}{2\pi} \frac{\tilde c_{ZZ}}{c_V}\sim {\mathcal O}(10^{-3})\,,
\end{equation}
for $\mathcal{O}(1)$ couplings. In the diphoton channel both terms are loop suppressed and the figure of merit for CP violation is
\begin{equation}
r_{\gamma\gamma}=  \frac{c \tilde c}{c^2 + \tilde c^2}\sim {\mathcal O}(1)\,,
\end{equation}
again assuming $\mathcal{O}(1)$ couplings. As will become clear, the measurement of CPV in $h\to \gamma\gamma$ is a challenging one, especially in comparison to the relatively straightforward measurement in $h\to ZZ^*$. However, the expected size of the effect may partially compensate for this. 

In addition to the CPV observables discussed in this paper, the CPV operator in Eq.~\eqref{eqn:LEFF} also modifies the overall $h\to \gamma\gamma$ decay rate, so that  
\begin{equation}
	\mu_{\gamma\gamma} \equiv \frac{\Gamma(h\to \gamma\gamma)}{\Gamma(h\to \gamma\gamma)_{\rm SM}} = \frac{c^2 + \ctilde^2}{c_{\rm SM}^2}~,
\end{equation}
where $\Gamma(h\to \gamma\gamma)_{\rm SM}$ is the SM rate.
The total rate is only quadratically sensitive to CP violating NP because the interference terms integrate to zero over phase space.In contrast, the differential rates contain CP odd terms proportional to $c  \ctilde$ and thus may be linear in the NP coupling. This can lead to substantial increase in sensitivity for small $\ctilde$. 

Before proceeding, two remarks are in order. First, it is important to mention that there are severe constraints on $y_e \ctilde $ from bounds on the electric dipole moment of the electron~\cite{McKeen:2012av,ACME:2013om,Fan:2013qn}.  Taking the electron yukawa, $y_e$, to be the SM one, this gives $\ctilde \lesssim 10^{-3}$. These bounds are, however,  absent in the limit where the 125 GeV Higgs does not couple to electrons (for other possibilities in concrete UV models, see~\cite{McKeen:2012av,Shu:2013uua}). A strong motivation for contemplating a non-zero value of $\ctilde $ is, for instance, that it would be generated by new CP sources in models that lead to electroweak baryogenesis, see e.g., \cite{Shu:2013uua}. An independent measurement of $\ctilde$ is thus desirable.

Second, we assumed here that $c$ and $\ctilde$ are real. In the SM $c$ obtains its dominant contribution from a $W$ loop and a smaller destructive contribution from a top quark loop. However, there is also a smaller contribution from $b$ quark and light quarks. These can go on-shell in the loop, generating complex effective couplings $c$ and $\ctilde$. This means that $c$ and $\ctilde$ obtain a relative strong phase. The result of such a strong phase, when combined with the weak phase, would be to induce direct CPV in decay, such that the decay rate of the Higgs into two positive helicity photons is not the same as into two negative ones. These strong phases are of $\sim{\mathcal O}( 1\%)$ in the SM \cite{Korchin:2013ifa,Korchin:2013jja}, and we assume the strong phases are similarly small for NP effects. Consequently, we neglect direct CPV, and assume that $c$ and $\ctilde$ are real.

\subsection{Helicity interference}
\label{sec:DRIE}
We proceed to examine the unconverted $\hgg$ process. The effective operator mediating $\hgg$ decay has the general form, cf. Eq.~\eqref{eqn:LEFF},
\begin{equation}
	\label{eqn:HEFF}
	\mathcal{H}_{\rm eff} = -c\frac{\alpha}{\pi v} h  F_{\mu\nu}F^{\mu\nu} - \frac{\ctilde}{2} \frac{\alpha}{\pi v} h  F_{\mu\nu} \tilde F^{\mu\nu}~.
\end{equation}
For a Higgs that is a scalar, the first term is CP even and the second is CP odd. CP is therefore violated if the CP phase 
\begin{equation}
	\xi\equiv \tan^{-1} (\ctilde /c)~,
\end{equation}
is found to be non-zero. 

The $\hgg$ helicity amplitudes are (dropping the overall $\alpha/\pi v$ factor, cf. Eq.~\eqref{eqn:HEFF})
\begin{align}
i\mathcal{M}^{\lambda_1\lambda_2} 
& = 
\parbox[c]{7cm}{
\includegraphics[scale=1]{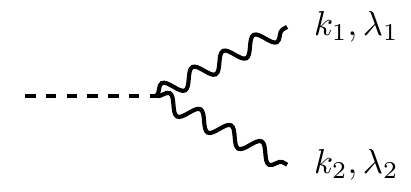} 
}\notag\\
& = c \Big[ (k_1\cdot k_2) ( (\varepsilon^{\lambda_1}_1)^* \cdot (\varepsilon^{\lambda_2}_2)^*) - (k_1\cdot   (\varepsilon^{\lambda_2}_2)^*) (k_2 \cdot  (\varepsilon^{\lambda_1}_1)^*)\Big] + \ctilde~ \epsilon\big[k_1 ,(\varepsilon^{\lambda_1}_1)^*, k_2, (\varepsilon^{\lambda_2}_2)^*\big]~, \label{eqn:HGGHA}
\end{align}
where $k_i$ are the photon momenta, $\varepsilon^{\lambda_i}_i$ is the polarization vector of the $i$th photon  ($i=1,2$) with helicity $\lambda_i = \pm$, and $\epsilon[p,q,r,s]\equiv p_\mu q_\nu r_\rho s_\sigma \epsilon^{\mu\nu\rho\sigma}$. A Latin subscript $i = 1,2$ hereafter denotes the corresponding photon.
To compute helicity amplitudes we employ the spinor-helicity formalism, see Appendix \ref{app:SHF} for our conventions and a brief review. Using Eqs.~\eqref{eqn:RMC}, \eqref{eqn:WSPC} and \eqref{eqn:ESI}, one finds that the non-zero helicity amplitudes are
\begin{equation}
	\label{eqn:HGGHAE}
	\mathcal{M}^{\pm\pm} = m_h^2(c \pm i\ctilde) = m_h^2 \sqrt{c^2 + \ctilde^2} \,e^{\pm i\xi}~,
\end{equation}
while $\mathcal{M}^{\pm\mp} = 0$ as expected from angular momentum conservation (cf. also the results of, e.g., \cite{Keung:2008ve,Cao:2009ah} for $h \to ZZ$). We see that CPV introduces a relative phase, $\xi$, between the two-photon helicity amplitudes. Furthermore, a differential rate may depend on $\xi$ if and only if there is interference between $\mathcal{M}^{++}$ and $\mathcal{M}^{--}$, or more precisely, between amplitudes involving the $++$ and $--$ photon helicity configurations. We call such interference \emph{helicity interference}. 

Let us now translate Eq.~\eqref{eqn:HGGHAE} into a Hilbert space language. The final states of $\hgg$ decay are the two-photon states $|\!+\!+\rangle$ and $|\!--\rangle$, with  $\pm$ indicating the helicity of each photon, so that CP$|\!\pm\!\pm\rangle = |\!\mp\!\mp\rangle$. Eq.~\eqref{eqn:HGGHAE} then translates to
\begin{align}
	\mathcal{H}_{\rm eff}|h\rangle 
	& \propto \sqrt{c^2 + \ctilde^2}\Big(e^{i\xi} \,|\!+\!+\rangle  + e^{-i\xi} \,|\!-\!-\rangle\Big)~. \label{eqn:HSHGG}
\end{align}
As above, the CP phase $\xi$ appears as a relative phase between the $|\!+\!+\rangle$ and $|\!-\!-\rangle$ terms. 

Now, the total rate for $\hgg$ decay is proportional to
\begin{equation}
	\sum_{f=\!+ +,\!--}\!\!\! |\langle f | \mathcal{H}_{\rm eff}|h\rangle|^2 = \sum_{f=\!+ +,\!--}\!\!\!|\mathcal{M}^f|^2~.
\end{equation}
Orthogonality of $|\!+\!+\rangle$ and $|\!--\rangle$ ensures that the total rate is independent of $\xi$, i.e., there is no helicity interference. In contrast, any experiment for which the final state is a linear superposition of the two helicity states would generate helicity interference. This is the case at collider experiments in which the on-shell photons with definite helicity are intermediate states: The final state is a converted photon -- an $e^+e^-$ pair with a particular set of momenta -- which has non-vanishing overlap with both helicities. 

\subsection{A thought experiment with polarizers}
\label{sec:ATE}

The overlap of each photon helicity with a BH pair will determine the strength of helicity interference and our ultimate sensitivity to the Higgs CP properties. The details at the level we need are quite involved, and will be described in Sec. \ref{sec:HBHP}. As a warm-up we instead consider a thought experiment in which we can measure linearly polarized photons.

 \begin{figure}[t]
\begin{center}
\includegraphics[width=5cm]{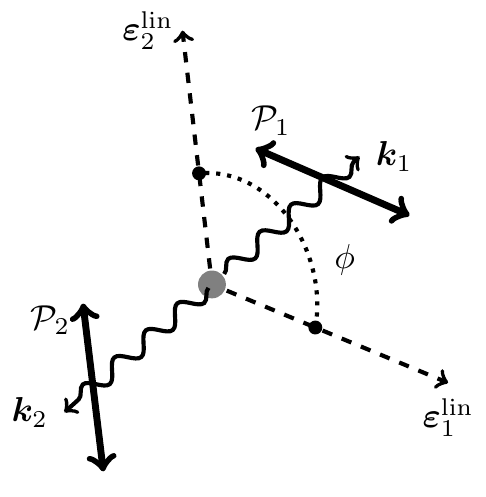}
\end{center}
\caption{A linear polarization thought experiment in Higgs rest frame. $\mathcal{P}_{1,2}$ are linear polarizers oriented orthogonal to the photon momentum direction. The angle $\phi$ is measured between the linear polarization vectors $\bm{\varepsilon}^{\rm lin}_{1,2}$.}
\label{fig:HGGE}
\end{figure}

Let us imagine that we have been able to manufacture a linear polarizer for gamma rays. We produce a Higgs at rest between polarizers $\mathcal{P}_1$ and $\mathcal{P}_2$, such that each photon travels through a polarizer (see Fig. \ref{fig:HGGE}) before being absorbed by a detector that counts photons. The polarizer $\mathcal{P}_i$ ($i=1,2$) projects an incoming photon onto a  linearly polarized state, $|\phi_i\rangle=e^{-i\phi_i} |+\rangle + e^{i\phi_i} |-\rangle$, that has polarization oriented at angle $\phi_i$. 
From Eq.~\eqref{eqn:HSHGG} the amplitude of the two-photon wave function observed by the detectors is
\begin{equation}
	\Big(e^{i\phi}\langle +| + e^{-i\phi}\langle - |\Big)_2\otimes\Big(\langle +| + \langle -|\Big)_1\mathcal{H}_{\rm eff}|h\rangle  \sim  \sqrt{c^2 + \ctilde^2}\cos(\phi + \xi)~, 
\end{equation}
where  $\phi=\phi_1-\phi_2$ is a relative azimuthal angle between the two polarizers. As the angle $\phi$ is changed, the differential rate in the detectors changes as $\cos^2(\phi+\xi)$. One finds
\begin{equation}
	\label{eqn:HGGDR}
 	\frac{d\Gamma}{d\phi} = \frac{2}{\pi}\Gamma_{\hgg} \cos^2(\phi + \xi)~,\qquad  \Gamma_{\hgg} = \frac{\alpha^2}{4\pi^3}\frac{m_h^3}{v^2}(c^2 + \ctilde^2)~.
\end{equation}
Note that the CP odd term in the differential rate \eqref{eqn:HGGDR} is  proportional to $\sin2\xi \sin2\phi\propto c\ctilde\sin 2\phi$. The differential rate is thus linearly sensitive to CPV coupling $\tilde c$, whereas the total rate is quadratically sensitive. 

In summary, we have shown that only the terms receiving contributions from both of the definite helicity two-photon amplitudes, so that there is helicity interference, are sensitive to the CPV phase $\xi$. These interference effects can in principle be of $\mathcal{O}(1)$ in size.

\section{Bethe-Heitler photon conversion}
\label{sec:BHPC}

We now study the process that can be used for photon polarization measurement, namely the conversion of a photon into an $e^+e^-$ pair in matter. In this section we study a conversion of single isolated photon, which we will then use in the Sec. \ref{sec:HBHP} for the case of $h\to \gamma\gamma$ with converted photons.

\subsection{Photon propagation and conversion}
\label{sec:PPC}
\begin{figure}[t]\centering
\includegraphics[scale=1]{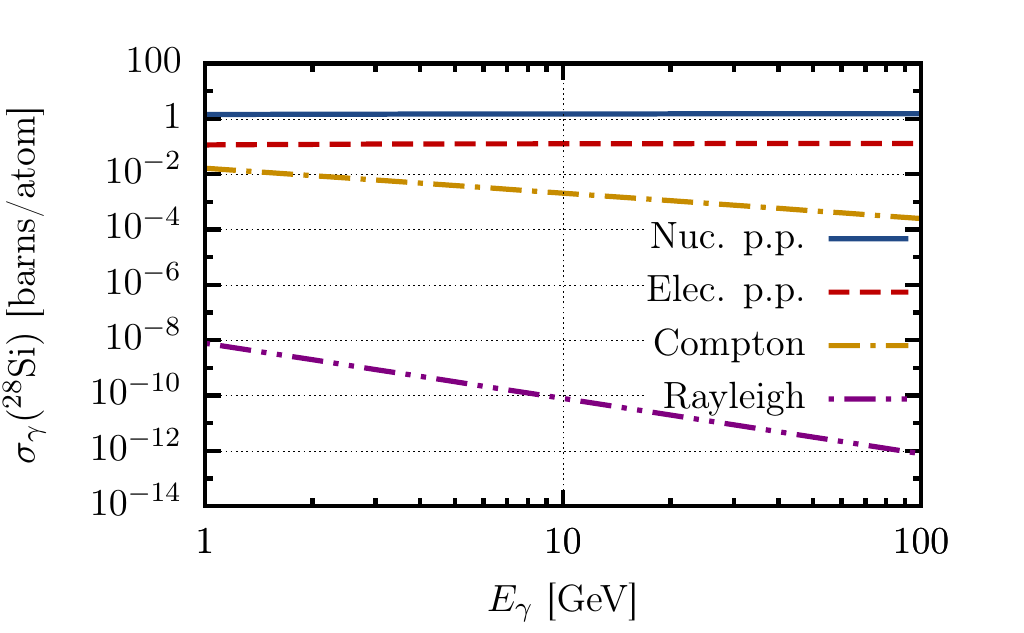} 
\caption{The contributions to photon cross-section on $^{28}\rm{Si}$, $\sigma_\gamma(^{28}\rm{Si})$, from BH $e^+e^-$ pair production in nuclear field (solid blue line), pair production due to scattering on electron cloud (red dashed), Compton scattering (dot-dashed yellow) and Rayleigh scattering (magenta double dot-dashed), as a function of photon energy $E_\gamma$. Calculated using NIST's XCOM database~\cite{berger2010xcom}.
}
\label{fig:photxs}
\end{figure}

Photon conversion to $e^+e^-$ pairs may proceed either by Dalitz conversion in vacuum, for an off-shell intermediate photon, or by BH conversion on atomic nuclei, which occurs for on-shell photons (for a review see, e.g., \cite{Tsai:1974py,Tsai:1977py}). The Dalitz conversion rate carries a suppression factor of ${\mathcal O}(10^{-4})$, and is not of immediate practical interest. Moreover it mainly proceeds via a longitudinal photon so that the above helicity analysis no longer applies. In contrast, the CMS and ATLAS pixel detectors contain a significant amount of material, so that $\sim 50$\% of photons convert inside the tracking systems via the BH process \cite{CMS-PAS-HIG-13-001,ATLAS-CONF-2013-012}. Based on the composition of the detectors, we assume in this work that the target nucleus is always $^{28}\rm{Si}$, at rest in the laboratory frame. This nucleus is spin-0, and has no nuclear magnetic moment. We therefore do not consider the effects of target polarization on the BH process \cite{Drell:1964ep}. 

One might be concerned by the prospect of photon polarization decoherence for the photons propagating  inside the silicon. However, at photon energies $\sim {\mathcal O}(m_h/2)$ the  pair production in the nuclear field is by far the largest contribution to the photon scattering cross-section in an atomic lattice \cite{pdg:2012}, see Fig.~\ref{fig:photxs}. As a result, to an excellent approximation, the photons remain coherent up until their BH conversion. We shall also assume that the BH scattering is quasi-elastic, i.e. that the target nucleus remains in a coherent state during and after the scattering. For the kinematics considered, the quasi-elastic limit is an excellent approximation of the full BH conversion \cite{Tsai:1974py,Tsai:1977py}.

\begin{figure}[t]\centering
\includegraphics[scale=1]{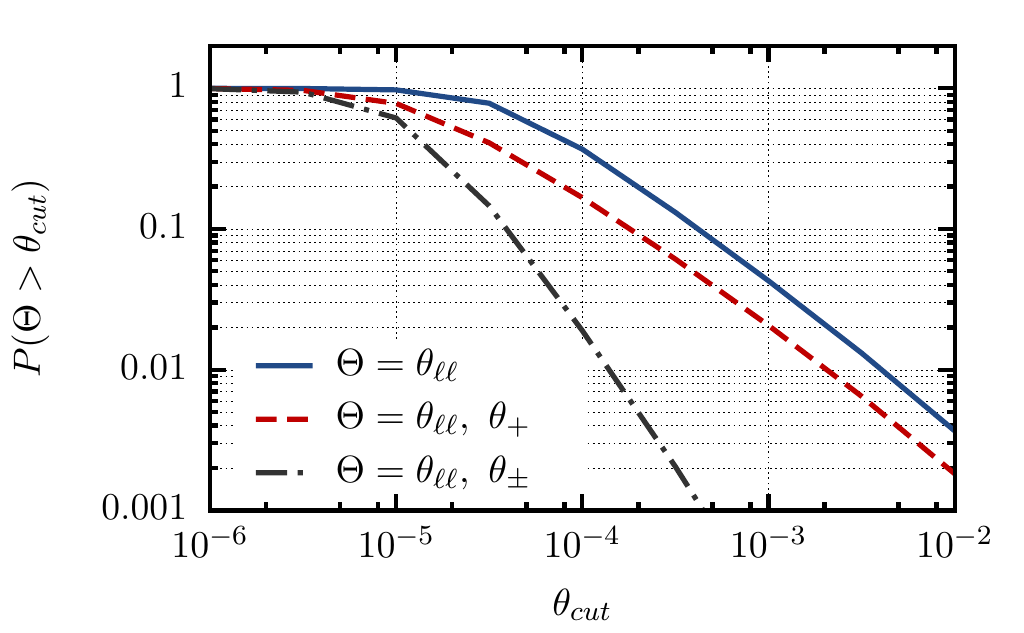} 
\caption{Cumulative distributions of Bethe-Heitler conversion cross-section for a photon with $E_\gamma=60$ [GeV] scattering on a $^{28}\rm{Si}$ nucleus, with respect to various opening angle cuts. Three distributions are shown: $P(\theta_{\ell\ell} > \theta_{\rm cut})$, i.e. with photon-lepton opening angles, $\theta_{\pm}$, unconstrained (blue line); $P(\theta_+ ~\mbox{and}~ \theta_{\ell\ell} > \theta_{\rm cut})$ with electron-photon opening angle $\theta_-$ unconstrained (red dashed line); $P(\theta_{\pm} ~\mbox{and}~ \theta_{\ell\ell} > \theta_{\rm cut})$ (black dot-dashed line).}
\label{fig:bhopang}
\end{figure}

\subsection{Angular resolution limitations}
\label{sec:AR}

Following the $\hgg$ experiment discussed in Sec. \ref{sec:ATE}, in order to measure the CPV in a doubly converted $h\to \gamma\gamma$ decay,  we might expect that angular distributions between planes formed by spatial momenta need to be measured. There are several possible planar distributions that can be constructed, involving either:
\begin{enumerate}
	\item The $e^+e^-$ plane formed by the $e^+e^-$ momenta, or;
	\item The $\gamma e^\pm$ plane formed by a lepton and its parent photon, as in Fig. \ref{fig:cartoon}.
\end{enumerate}
The first requires resolving the orientation of the leptonic spatial momenta. The second requires the orientation of the leptons with respect to their parent photon, which could be achieved by identifying the vertex associated with the Higgs (from other tracks in the event) as well as the location of the photon conversion, giving the photon direction. 

Such measurements require exquisitely precise tracking. Because the momentum transfer to the nucleus is small, the relative angular orientations between the photon and leptons are tiny in typical photon conversion: for mass $m$ and energy $E$, the angular scale is typically $m/E \sim 10^{-5}$ for a $60$ GeV photon conversion to electrons. There is however a distribution for these angles. In the limit of very large statistics one can hope to get a sample of events where these angles can be measured.

In the ATLAS detector, for instance, the intrinsic accuracy in silicon pixels located between 5cm and 12 cm from the interaction point is $10\mu m$ in $R-\phi$ direction and $115\mu m$ in $z$ direction. The intrinsic accuracy of SCT strips located between 30cm to 51cm from the interactions point is $17 \mu m$ ($R-\phi$) and $580\mu m (z)$ \cite{Aad:2008zzm}. One may therefore hope to measure the orientations of the $e^+e^-$ plane even for opening angles as small as $\theta_{\ell\ell} \sim 10^{-4}-10^{-3}$, where the relative leptonic angle $\theta_{\ell\ell}$ is defined by
\begin{equation}
	\label{eqn:DTLL}
	\cos \theta_{\ell\ell} = \frac{\bm{p}_+ \cdot \bm{p}_-}{|\bm{p}_+||\bm{p}_-|}~,
\end{equation}
for leptonic spatial momenta $\bm{p}_\pm$. By comparison, a 60 GeV photon converting to a $e^-e^+$ pair has an opening angle of $\theta_{\ell\ell} >10^{-4}$ in $38\%$ of the cases and $\theta_{\ell\ell}>10^{-3}$ in $4\%$ of the cases. The full cumulative distribution for  $\theta_{\ell\ell} > \theta_{\rm cut}$, is shown in Fig.~\ref{fig:bhopang}.

Another experimental challenge is the multiple scattering of outgoing electrons when traversing the detector medium. This can affect the measurement of the electron direction and thus the orientation of the $e^+e^-$ or $\gamma e^\pm$ plane. Using Eq.~(30.15) in \cite{pdg:2012}, the width of the angular distribution is $\sim 10^{-4}$ for a $30$ GeV electron, assuming it traverses $\ell = 0.1$ radiation lengths of the material. This width roughly scales as $\sqrt{\ell}/E$, where $E$ is the lepton energy. The measurement of polarization planes in the current and future detectors will thus be challenging, but may be achievable on a statistical basis. Bearing in mind these experimental questions, in the remainder of this paper we adopt a theoretical approach to this problem: We consider a thought experiment where all angles can be resolved and explore the sensitivity to Higgs parameters in this best-case scenario. 

\subsection{Nuclear form factor}
\begin{figure}[t]\centering
	\includegraphics[scale=1]{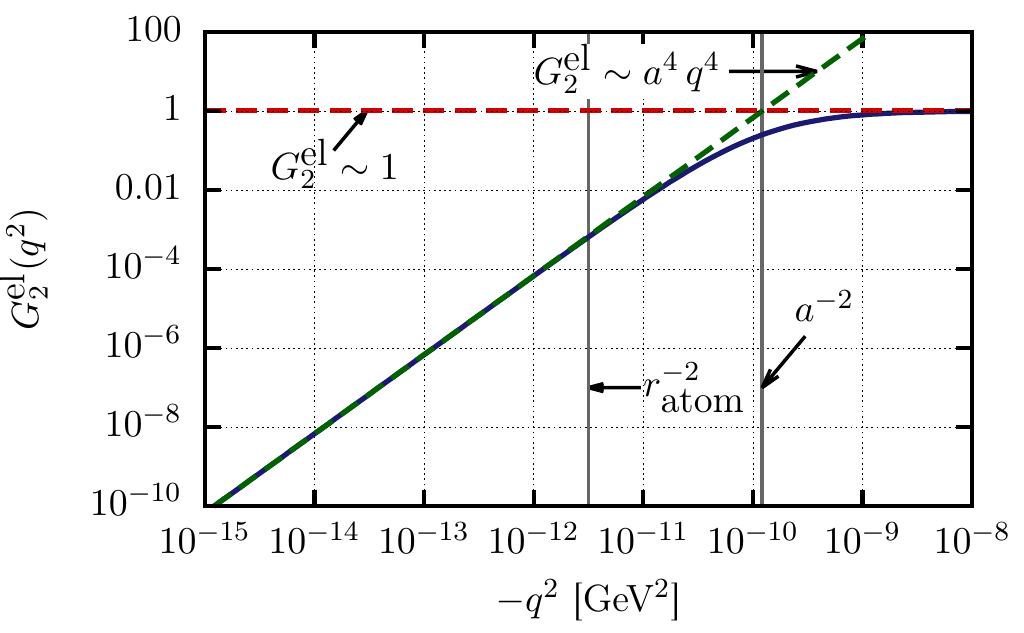}
\caption{The elastic form factor $G_2^{\rm el}(q^2)$. The dashed lines show the limiting behavior $G_2^{\rm eq} \sim a^4 q^4$ for $|a^2\,q^2|\ll 1$ (green dashed line) and $G_2^{\rm el} \sim 1$ for $|a^2\,q^2|\gg 1$ (red dashed line). The scale at which screening of the nucleus becomes important is denoted by $a$, which is smaller than the Si atomic radius, $r_\text{atom}$. At scales well outside the atom, corresponding to small $-q^2$, nuclear conversion is suppressed by the form factor screening. 
}
\label{fig:FF}
\end{figure}

The BH conversion depends on a momentum transfer, $q^\mu$,  between the photon and the nucleus. Assuming quasi-elastic scattering, the photo-nuclear scattering is encoded in an elastic nuclear form factor $G_2^{\rm el}[q^2]$ (see Eq.~\eqref{eqn:GFFD} below). This form factor plays an important role of suppressing scattering at low $-q^2$, that is, at scales larger than the Si atom.  

Let us discuss briefly  the behavior of the nuclear form factor. 
The threshold for an $E_\gamma \simeq 60$ GeV photon to convert to an $e^+e^-$ pair is at $-q^2=4m^4/E_\gamma^2\simeq 10^{-18}$ GeV$^2$, with  $m$ the electronic mass, but can occur at momentum transfers as large as $-q^2 \sim 10 ^{-6}$ GeV$^2$. This should be compared with the radius of the Si atom, $r_{\rm atom}\simeq 1.1{\textrm \AA}$~\cite{1964JChPh..41.3199S} or  $r_\text{atom}^{-2}\simeq 10^{-12}$ GeV$^2$, and with the nuclear radius $r_{\rm nuclear}\simeq 4$ fm which gives $r_\text{nucl}^{-2}\simeq 10^{-3}$ GeV$^2$. 
Within the $-q^2$ range of interest for conversion --  $10^{-18}$ up to $10^{-6}$ GeV$^2$ -- the nuclear charge is thus screened at low $-q^2$ by the atomic shell electrons. In this work we use the simple expression for the atomic form factor~\cite{schiff1951energy}  described in detail in~\cite{Tsai:1974py}  and given by
\begin{equation}\label{G2el}
	G_2^{\rm el}(q^2)= \frac{M^2 a^4 q^4}{(1- a^2 q^2)^2}~,
\end{equation}
where $a = 184.15(2.718)^{-1/2}Z^{-1/3}/m$ and $Z$ is the atomic number of the nucleus with mass $M$. For $^{28}$Si, $a^{-2} = 1.22 \times 10^{-10}$ GeV$^2$. 
There are two limits of interest. The first is $|a^2\,q^2|\gg 1$ in which $G_2^{\rm el} \sim 1$, the second is the limit $|a^2\,q^2|\ll 1$ in which $G_2^{\rm el} \sim q^4$, see  Fig. \ref{fig:FF}. That is,  the form factor suppresses the BH cross-section for small $-q^2 \ll a^2$. To the extent that the $1/q^4$ factor in the BH cross-section determines the dominant phase space configurations of the final states, this suppression significantly alters the important regions of phase space for BH conversion up to the $a^{-2}$ scale. Specifically, the form factor increases the probability of significantly acoplanar BH conversions.

\section{The Higgs-Bethe-Heitler process}
\label{sec:HBHP}
In this section we present a formal analysis of the Higgs-Bethe-Heitler (HBH) process, $h\to \gamma\gamma$ with both photons converting to $e^+e^-$ pairs. The main result is that we obtain compact, leading order 
expressions for the HBH rate, and gain insight into the structure of the terms sensitive to CP violation.

\subsection{Amplitude and cross-section}
\label{sec:ACS}

The amplitude of interest is given by a menorah diagram, consisting of a $\hgg$ and two BH conversion subdiagrams, summed over internal photon polarizations, viz.
\begin{align}
i\mathcal{M}^{\mu\nu}_{1_\rs 2_\rs} &=
\parbox[c]{9cm}{
\includegraphics[width=7cm]{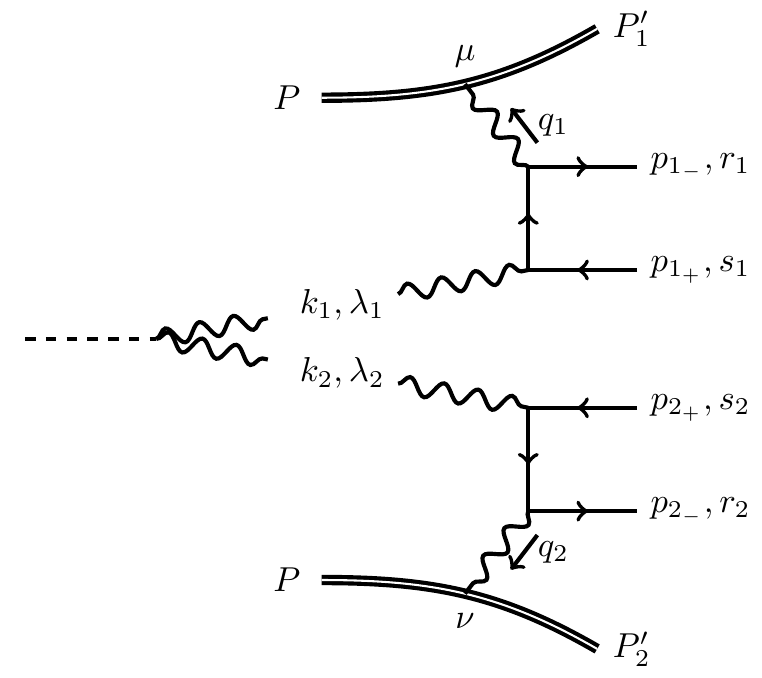}
} + \mbox{lepton exchanges}~\notag\\
& = \frac{1}{\Lambda}\sum_{\lambda_1,\lambda_2}\Big[ c\Big( k_1\cdot k_2 g^{\alpha\beta} - k_1^\beta k_2^\alpha\Big) + \tilde{c} \Big(k_{1_\rho}k_{2_\sigma}\epsilon^{\rho\alpha \sigma \beta}\Big)\Big] (\varepsilon_\alpha^{\lambda_1})^*(\varepsilon_\beta^{\lambda_2})^*
[{M^\mu_{\rm BH}}]^{\lambda_1}_{1_\rs}[{M^\nu_{\rm BH}}]^{\lambda_2}_{2_\rs}~, \label{eqn:HBH}
\end{align}
in which $\Lambda = \pi v/e^6\alpha$, for QED coupling $e$. The BH amplitudes are
\begin{equation}
\label{eqn:FFSA}
[{M^\mu_{\rm BH}}]^{\lambda_i}_{i_\rs} =  \bar{u}^{r_i}(\Pm i)\bigg[\slashed{\varepsilon}^{\lambda_i} \frac{i}{\slashed{p}_{i_-} - \slashed{k}_i -m} \gamma ^\mu + \gamma ^\mu\frac{i}{ \slashed{k}_i -\slashed{p}_{i_+} -m}\slashed{\varepsilon}^{\lambda_i}\bigg] v^{s_i}(\Pp i)~,
\end{equation}
where we have not yet taken nuclear form factors into account, and kept explicit the corresponding Lorentz index of the nuclear electromagnetic current. The Latin subscripts, $i=1,2$, label each photon, while $s_i$, $r_i = 1,2$ are respectively the positron and electron spins (see Eq.~\eqref{eqn:SLCD}), and $\lambda_i = \pm$ the outgoing photon helicities from the $\hgg$ vertex. We have also suppressed repetitions of the photon index, such that $X_{i_\rs}$ is henceforth a shorthand for $X_{i_{r_i s_i}}$. We shall often refer to the BH subdiagrams for each photon as the photon branch. 

We assume both nuclei are initially at rest in the lab frame, so $P^\mu = (M,\bm{0})$ where~$M$ is the mass of the nucleus. The Higgs need not be at rest in the lab frame. As discussed in Sec. \ref{sec:PPC}, we assume quasi-elastic scattering, that is $P_i'^2 = P^2 = M^2$. This implies that $2q_i\cdot P + q_i^2 =0$, and so
\begin{equation}
	E_i - E_{i_+} - E_{i_-} + q_i^2/2M  = 0~,\label{eqn:QSA}
\end{equation}
where $E_i$, $E_{i_\pm}$ are the photon and lepton energies respectively. It is also convenient to define
\begin{equation}
Q^\mu_i \equiv \frac{1}{M}\bigg(P^\mu - q_{i}^\mu\frac{ q_{i}\cdot P}{q_i^2}\bigg)  = \frac{1}{M}\bigg(P^\mu + \frac{q^\mu_i}{2}\bigg)~,\\
\end{equation}
under quasi-scattering conditions.  The exchange energy with the nucleus $q_i^0 \equiv P^{'0}_i - P^0_i \ll M$. I.e. the nucleus velocity is non-relativistic, so to an excellent approximation, it follows that
\begin{gather}
	\label{eqn:EQA}
	E_i \simeq E_{i_+} + E_{i_-}~, \qquad Q_i^\mu \simeq (1, \bm{0})~,
\end{gather}
in the lab frame.

We define the BH nuclear form factor tensors on each photon branch \cite{Tsai:1974py},
\begin{equation}
	\label{eqn:WST}
	\mathcal{W}_{i}^{\mu\nu} = - W_1(q_i^2) \bigg(g^{\mu\nu} - \frac{q_{i}^{\mu} q_{i}^{\nu}}{q_i^2}\bigg)+ W_2(q_i^2) Q_i^{\mu} Q_i^\nu~,
\end{equation}
such that the unpolarized HBH squared amplitude 
\begin{equation}
	\label{eqn:FFGD}
	|\mathcal{M}|^2 \prod_{i=1,2} 2 M\delta(M^2-P_i'{}^2)=  \frac{1}{q_1^4q_2^4}\sum_{r_i,s_i} \mathcal{M}^{\mu\nu}_{1_\rs 2_\rs}\mathcal{M}_{1_\rs 2_\rs}^{*\mu'\nu'}{\mathcal{W}_1}_{\mu\mu'}{\mathcal{W}_2}_{\nu\nu'}~.
\end{equation}
On the left of Eq.~\eqref{eqn:FFGD} we have factored out the $\delta-$functions that enforce quasi-elastic scattering. The form factor $W_1(q^2) = 0$ for quasi-elastic scattering on a spin-0 nuclear target \cite{Drell:1964ep}, while 
\begin{equation}
	\label{eqn:GFFD}
	W_2(q^2)=2 M \delta(M^2-P'{}^2) G_2^{\rm el}(q^2)~,
\end{equation}
in which $G_2^{\rm el}(q^2)$ is the form factor given in Eq.~\eqref{G2el}. Hence the unpolarized HBH squared amplitude reduces to
\begin{equation}
	\label{eqn:HBHUR}
	|\mathcal{M}|^2 =  \frac{G_2^{\rm el}(q_1^2)G_2^{\rm el}(q_2^2)}{q_1^4q_2^4}\sum_{r_i,s_i} \mathcal{M}^{\mu\nu}_{1_\rs 2_\rs}\mathcal{M}_{1_\rs 2_\rs}^{*\mu'\nu'} {Q_1}_{\mu}{Q_1}_{\mu'}{Q_2}_{\nu}{Q_2}_{\nu'}~.
\end{equation}

\subsection{Helicity structure}\label{sec:hel}
One may calculate $|\mathcal{M}|^2$ via the usual Feynman method, which relies on polarization completeness relations to compute traces. This approach leads to thousands of terms, of which the na\"\i vely dominant terms cancel due to Ward identities. Extracting leading order expressions is therefore difficult, and moreover, high numerical precision is required for numerical stability. As an alternative, we employ a spin and helicity analysis combined with the spinor-helicity formalism to compute the HBH amplitudes. These may be subsequently squared and summed over external spins to produce the full HBH rate. In the following we provide a brief overview of these results, while details are provided in Appendices \ref{app:SHF} and \ref{app:SHA}.

The BH spin-helicity  amplitudes are defined as
\begin{equation}
	\label{eqn:MQA}
	\alpha_{i_\rs}^\pm \equiv [M^\mu_{\rm BH}]^\pm_{i_\rs}{Q_i}_\mu/q_i^2~.
\end{equation}
With reference to Eqs.~\eqref{eqn:HBH} and \eqref{eqn:HBHUR}, the HBH spin-helicity amplitudes are correspondingly
\begin{equation}
i\mathcal{M}^{\lambda_1\lambda_2}_{1_\rs 2_\rs}  = \frac{ \sqrt{\mcG_1\mcG_2}}{\Lambda}\Big[ c\Big( k_1\cdot k_2 g^{\alpha\beta} - k_1^\beta k_2^\alpha\Big) + \tilde{c} \Big(k_{1_\rho}k_{2_\sigma}\epsilon^{\rho \alpha\sigma \beta}\Big)\Big] (\varepsilon_\alpha^{\lambda_1})^*(\varepsilon_\beta^{\lambda_2})^*\alpha_{1_\rs}^{\lambda_1}\alpha_{2_\rs}^{\lambda_2}~,
\end{equation}
where we introduced the abbreviation
\beq
\mcG_i \equiv G_2^{\rm el}(q_i^2)~.
\eeq
Making use of the spinor-helicity formalism (see Appendix \ref{app:SHF}), we obtain the spin amplitudes
\begin{equation}
	\mathcal{M}_{1_\rs2_\rs} \equiv
	\sum_{\lambda_1\lambda_2}\mathcal{M}^{\lambda_1\lambda_2}_{1_\rs 2_\rs} = 
	\frac{m_h^2}{\Lambda} \sqrt{\mcG_1\mcG_2}\sqrt{c^2 + \ctilde^2} 
	\Big(e^{i\xi}\alpha^+_{1_\rs}\alpha^+_{2_\rs} +  e^{-i\xi}\alpha^-_{1_\rs}\alpha^-_{2_\rs}\Big)~.
\end{equation}
These are reminiscent of equation~\eqref{eqn:HSHGG} with final leptonic spin states $\langle r_i s_i |$ and $\alpha^\pm_{i_\rs} \propto \langle s_i r_i | \pm \rangle$, as expected.
The HBH square-amplitude is correspondingly
\begin{equation}
	\label{eqn:HBHA}
	|\mathcal{M}|^2 \equiv  \sum_{1_\rs 2_\rs}\big|\sum_{\lambda_1\lambda_2}\mathcal{M}^{\lambda_1\lambda_2}_{1_\rs 2_\rs}\big|^2=m_h^4\mcG_1\mcG_2 \frac{c^2 + \tilde{c}^2}{4\Lambda^2}\sum_{1_\rs 2_\rs}\Big|\alpha^+_{1_\rs}\alpha^+_{2_\rs} e^{i\xi} + \alpha^-_{1_\rs}\alpha^-_{2_\rs} e^{-i \xi}\Big|^2~.
\end{equation}
Note that the sum over photon polarizations is inside the absolute value as expected for entangled $h\to 2\gamma\to 2(e^+e^-)$ decay, but the sum over lepton spins is incoherent. Eq.~(\ref{eqn:HBHA}) shows 
that the BH spin-helicity amplitudes $\alpha^\pm_{i_\rs}$ are all one needs to determine the entire HBH square amplitude. Parity invariance of the BH amplitudes relates amplitudes of opposite helicity and spins to their complex conjugates,
\begin{equation}
	\label{eqn:PA}
	(\alpha^+_{rs})^* = \eta_{rs}\alpha^-_{\bar{r}\bar{s}}~,
\end{equation}
where $\bar{s}$ is the opposite spin to $s$, and $\eta_{rs}^2 = 1$, $\eta_{rs} = \eta_{\bar{r}\bar{s}}$. Hence we need only determine $\alpha^+_{i_\rs}$. We shall see below that spinor-helicity methods, when applied to $\alpha^\pm_{i_\rs}$, also allow for a well-controlled expansion of dominant, sub-dominant and negligible terms in $|\mathcal{M}|^2$. 

We see explicitly from Eq.~\eqref{eqn:HBHA} that the helicity interference terms are equivalent to the $\xi$-dependent terms, as discussed in sections~\ref{sec:DRIE} and~\ref{sec:ATE}. For a particular leptonic spin configuration $\{r_i,s_i\}$, helicity interference occurs so long as $\alpha^{+}_{i_\rs}$ and $\alpha^{-}_{i_\rs}$ are both non-zero. However, we see in appendix \ref{app:SHA}, and in particular in Eq.~\eqref{eqn:BHSALO}, that for our particular choice of spinor basis (see Eq.~\eqref{eqn:SLCD})
\begin{equation}
	\label{eqn:AH}
	 |\alpha^-_{i_{11}}| \sim |\alpha^-_{i_{12}}| \sim |\alpha^-_{i_{21}}| \gg |\alpha^-_{i_{22}}|~,
\end{equation}
or equivalently $ |\alpha^+_{i_{22}}| \sim |\alpha^+_{i_{12}}| \sim |\alpha^+_{i_{21}}| \gg |\alpha^+_{i_{11}}|$. This hierarchy means that we may therefore discard any terms containing either $\alpha^+_{i_{22}}$ or $\alpha^-_{i_{11}}$ as subleading. It follows that the leading order squared amplitude is
\begin{equation}
	|\mathcal{M}|^2 
	 \simeq m_h^4 \frac{c^2 + \tilde{c}^2}{4\Lambda^2} \mcG_1\mcG_2 \Bigg\{ \! 2\Big|\alpha^-_{1_{11}}\alpha^-_{2_{11}}\Big|^2  \! +2 \! \mathop{\mathop{\sum}_{j \not= k}}_{r_{k}\not=s_{k}}\!\!\Big|\alpha^-_{j_{11}}\alpha^-_{k_{rs}}\Big|^2 \! + \! \mathop{\mathop{\sum}_{r_1\not=s_1}}_{r_2\not=s_2}\!\!\Big|\alpha^+_{1_\rs}\alpha^+_{2_\rs}  e^{i\xi} + \alpha^-_{1_\rs}\alpha^-_{2_\rs} e^{-i \xi} \Big|^2\!\Bigg\},\label{eqn:HBHAE}
 \end{equation}
 while the leading interference term is
\begin{equation}
	|\mathcal{M}|^2\big|_{\rm int}  \simeq 2m_h^4\frac{c^2 + \tilde{c}^2}{\Lambda^2} \mcG_1\mcG_2
	\mbox{Re}\Big\{ \alpha_{1_{12}}^-\alpha_{1_{21}}^-\alpha_{2_{12}}^-\alpha_{2_{21}}^-e^{-2i\xi}\Big\}~.\label{eqn:CPOE}
\end{equation}
We see that only $\alpha^-_{12,21}$ (or equivalently $\alpha^+_{21,12}$) amplitudes enter the leading order $\xi$-dependent interference terms. 

The CP odd helicity interference term in \eqref{eqn:CPOE} is proportional to $c\ctilde$,
\begin{equation}
	\label{eqn:CPO}
	|\mathcal{M}|^2\big|_{\rm int, CP-odd} = -2m_h^4 \frac{c\tilde{c}}{~\Lambda^2} \mcG_1\mcG_2 \mbox{Im}\Big\{ \alpha_{1_{12}}^-\alpha_{1_{21}}^-\alpha_{2_{12}}^-\alpha_{2_{21}}^-\Big\}~.
\end{equation}
Note that interference terms between amplitudes, produced by the $FF$ and $F\tilde{F}$ operators respectively, are CP odd. However, the helicity amplitudes under consideration here receive contributions from both CP odd and CP even operators -- manifestly they depend on $\xi$ -- and hence \emph{helicity interference} terms contain both CP-even and CP-odd pieces. Consequently, we interpret the remaining piece of the helicity interference term to be the CP-even piece,
\begin{equation}
	\label{eqn:CPE}
	|\mathcal{M}|^2\big|_{\rm int, CP-even} = m_h^4 \frac{c^2 - \tilde{c}^2}{\Lambda^2} \mcG_1\mcG_2 \mbox{Re}\Big\{\alpha_{1_{12}}^-\alpha_{1_{21}}^-\alpha_{2_{12}}^-\alpha_{2_{21}}^-\Big\}~.
\end{equation}
This term has quadratic dependence on $\tilde c^2$, albeit a different one than the total rate \eqref{eqn:HGGDR}.

\subsection{The Bethe-Heitler helicity amplitudes}

In Eqs.~\eqref{eqn:HBHAE} and \eqref{eqn:CPOE} we have expressed the leading order HBH rate and interference term in terms of individual BH helicity amplitudes, $\alpha^\pm_{rs}$. In this subsection we proceed to present the leading order terms of these amplitudes in a readily accessible notation, for the special case that the Higgs is at rest in the lab frame. The results below are achieved via spinor-helicity techniques; a more comprehensive presentation of the derivation of these results is provided in Appendix \ref{app:SHA}. There, explicit results for each spin helicity amplitude $\alpha^\pm_{i_\rs}$ are collected in Eqs.~\eqref{eqn:BHSA}, while the leading order results, obtained by power counting in $m/m_h$, are provided in \eqref{eqn:BHSALO}. The compact results below will permit us, in the next section of this paper, to study the encoding of the CPV structure in the HBH rate at an analytic level. 

Before proceeding, we may first derive a new result concerning the well-studied unpolarized BH square amplitude, $|\mathcal{M}_{{\rm BH}_i}|^2 = \sum_{rs\lambda}|\alpha^\lambda_{i_\rs}|^2$.  Using Eqs.~\eqref{eqn:BHSA} one may show that
\begin{equation}
	\label{eqn:BHLOR}
	|\mathcal{M}_{{\rm BH}_i}|^2 \simeq 8\frac{\mcG_i m^2}{q_i^4} \bigg[\frac{E_{i_-} (k_i \cdot p_{i_-}) - E_{i_+} (k_i \cdot p_{i_+})}{(k_i\cdot p_{i_-}) (k_i \cdot p_{i_+})}\bigg]^2- 4\frac{\mcG_i}{q_i^2}\frac{ E_{i_+}^2 + E_{i_-}^2}{(k_i\cdot p_{i_-}) (k_i \cdot p_{i_+})}~.
\end{equation}
This compact expression for the BH rate in the quasi-elastic scattering limit is novel to this work.  A numerical analysis and validation of the resulting BH differential cross-section is provided in Appendix  \ref{app:BHDR}.

\begin{figure}[t]\centering
\includegraphics[scale=1.1]{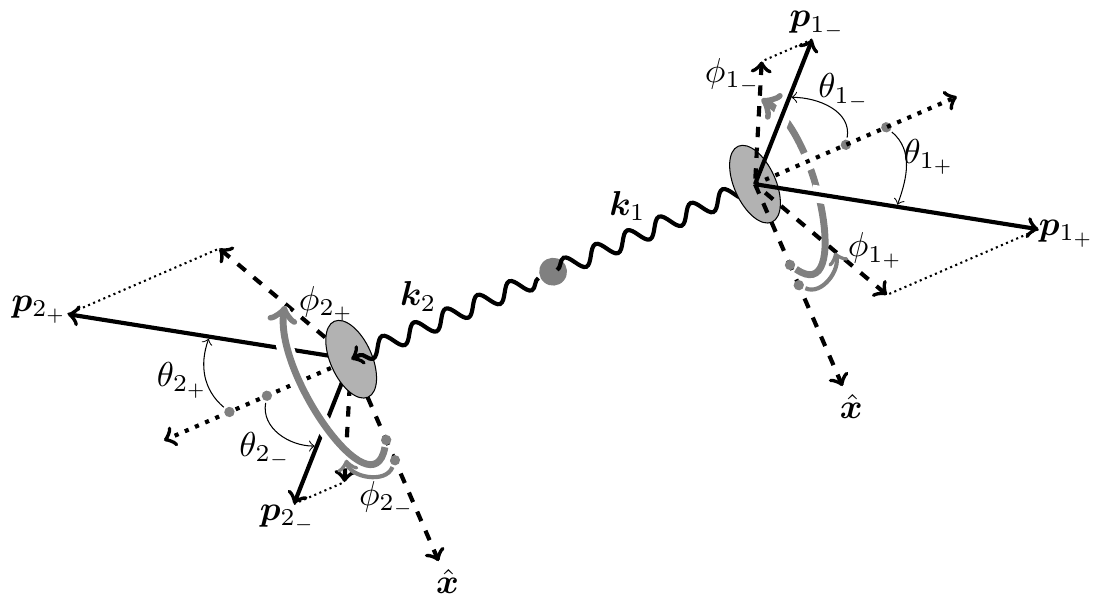}
\caption{Definitions of local spherical polar angles $\{\theta_{i_{\pm}}, \phi_{i_{\pm}}\}$. Note in particular that azimuthal angles, $\phi_{i_{\pm}}$ are positively oriented with respect to their parent photons, and are defined with respect to an azimuth $\hat{\bm{x}}$ common to both branches. Polar angles are defined with respect to parent photon momentum (black dotted). 
}
\label{fig:PAD}
\end{figure}

Now, in the special case that the Higgs is at rest in the lab frame, that is $\bm{P}_h = \bm{0}$, Eqs.~\eqref{eqn:BHSALO} for the $\alpha^\pm_{i_\rs}$ collapse to very simple expressions. Using the leading order results \eqref{eqn:BHSALO} and assuming  $\theta_\pm \ll 1$, for each branch one finds
\begin{equation}
\begin{split}
	\alpha^-_{11}  = - (\alpha^+_{22})^* & \simeq \frac{2 \sqrt{2\gamma_+\gamma_-}}{q^2} \bigg( \frac{1}{1 + \gamma_+^2\theta_+^2} - \frac{1}{1 + \gamma_-^2\theta_-^2}\bigg),\\
	\alpha^-_{^{12}_{21}}  =  +(\alpha^+_{^{21}_{12}})^* &  \simeq  \pm \frac{2 \sqrt{2\gamma_+\gamma_-}}{q^2} \frac{\gamma_{\mp}}{\gamma_+ + \gamma_-} \bigg( \frac{\gamma_+\theta_+ e^{-i\phi_+} }{1 + \gamma_+^2\theta_+^2} +  \frac{\gamma_-\theta_- e^{-i\phi_-} }{1 + \gamma_-^2\theta_-^2}\bigg),\\
	\alpha^-_{22}  = -(\alpha^+_{11})^* &  \simeq 0 ~.\label{eqn:BHSALOA}
\end{split}
\end{equation}
Here $\theta_\pm$ and $\phi_\pm$ are respectively the polar and azimuthal angles defined with respect to the branch parent photon momentum, as shown in Fig.~\ref{fig:PAD}, and $\gamma_\pm \equiv E_\pm/m \gg 1$. We also assume that the $\mathcal{O}(\theta)$ terms do not cancel completely. The latter may occur on the phase space slice $\gamma_+\theta_+ = \gamma_-\theta_-$ and $|\phi_- - \phi_+| =  \pi$, corresponding to minimal $|q^2|$.  Excellent numerical validation of this expansion implies that the neighborhood of this phase space slice, on which the expansion fails, is actually of negligibly small measure. Finally, we may approximate $q^2$ by 
\begin{equation}
	\label{eqn:QPAE}
	-q^2  \simeq m^2\Big(\gamma_+^2\theta_+^2 + \gamma_-^2\theta_-^2 + 2\gamma_-\gamma_+\theta_-\theta_+\cos(\phi_- - \phi_+)\Big) + \frac{m^2}{4}\bigg[\frac{1}{\gamma_+} + \frac{1}{\gamma_-}\bigg]^2~.
\end{equation}

Eqs.~\eqref{eqn:BHSALOA} and \eqref{eqn:QPAE}, when combined with Eq.~\eqref{eqn:HBHA}, form one of the central results of this paper: a compact form of the HBH rate, which is both numerically stable, and whose structure may now be studied analytically. For example,  it is now manifest that HBH polar angular structure is dominantly controlled just by the $k\cdot p_+/m^2 \simeq 1+ \gamma_\pm\theta_\pm$ denominators, which produce a peak near $\theta_\pm\gamma_\pm \sim 1$. More importantly, we see that the $\alpha^{\pm}_{12,21}$ amplitudes, that control the helicity interference terms, contain phases which are the leptonic azimuthal orientations, $\phi_{i_\pm}$. This non-trivial result, when combined with Eq.~\eqref{eqn:CPOE} shows us that, at leading order, $\xi$ manifests as a phase shift in the relative azimuthal orientations of leptons with different parent photons.

\section{Sensitivity to CPV}
\label{sec:O}

In this section we assess the potential sensitivity to CP violation. To do this, we consider a number of CPV sensitive observables and propose several sets of kinematic cuts that can enhance the CPV signal. These sets of cuts require that the lepton-lepton opening angle \eqref{eqn:DTLL}, $\theta_{\ell\ell}$, can be resolved, as well as the two photon-lepton angles, $\theta_\pm$. In the following we mostly consider angular resolution cuts of the form 
\begin{equation}
	\label{eqn:ARC}
	\theta_{\ell\ell}~, ~~\theta_{\pm} > \theta_{\rm cut}~.
\end{equation}
Following Sec. \ref{sec:AR}, we apply an angular resolution cut $\theta_{\rm cut}=10^{-4}$, which is at the edge of what may be possible with present detectors, and a looser, futuristic $\theta_{\rm cut}=10^{-5}$, intended to show that very large CPV effects are possible in principle.

We emphasize that as our measurement is a novel, challenging one, our goal here is not to conduct a full collider analysis including backgrounds. Rather our aim is to identify the types of observables that can probe CP violation in $h\to\gamma\gamma$, and estimate how well they do under ideal circumstances: a high efficiency in reconstructing conversions and a signal rich channel. This is in the anticipation that such circumstances might materialize in a future LHC running or at a Higgs factory. 

\subsection{Differential scattering rate}
The HBH differential scattering rate for the full $3 \to 6$  process (Higgs plus two nuclei to two nuclei and two $e^+e^-$ pairs) in the lab frame is
\begin{align}
	d \Gamma 
	& \propto  |\mathcal{M}|^2 d\Pi_{\hgg} d\Pi_{{\rm BH}_1} d\Pi_{{\rm BH}_2} \notag\\
	& \propto   \frac{|\mathcal{M}|^2}{(2\pi)^{12}M^2 E_\gamma^2} \bigg[\mathop{\prod_{i = 1,2}}_{\alpha = \pm} \frac{d^3\bm{p}_{i_\alpha}}{2E_{i_\alpha}}\bigg]\bigg[\prod_{i = 1,2} \frac{d^3\bm{P}^\prime_i}{2E^\prime_i} \frac{ d^3 \bm{k}_i}{2E_{i}}\delta^{(4)}(q_i+ p_{i_+} \!+ p_{i_-} \!- k_i) \bigg]\delta^{(4)}(P_h - k_1 - k_2)~,
\end{align}
with $|\mathcal{M}|^2$ given by Eq.~\eqref{eqn:HBHA}. Integrating over the out-going nuclear momenta and all other momenta in delta functions, 
we obtain in the limit $|q| \ll M$
\begin{equation}	
	d\Gamma \propto   \frac{1}{(2\pi)^{12}}\frac{m_h}{4 M^4 E_\gamma^2} |\mathcal{M}|^2 \bigg[\mathop{\prod_{i = 1,2}}_{\alpha = \pm} |\bm{p}_{i_\alpha}| d\Omega_{i_\alpha}\bigg]dE_{1_-}dE_{2_-}d\Omega_1~,
\end{equation}
with $d\Omega_{i_\alpha}$ the solid angle for each lepton momentum $\bm{p}_{i_\alpha}$, and $d\Omega_1$ the solid angle for photon~`1'. 
The photon labels are extrinsic. We take photon `1', say,  as westwards going (if Higgs is not at rest one can also take it to be, e.g., the more energetic photon).

For simplicity, we assume henceforth that the Higgs is at rest in the lab frame, that is $P_h = (m_h,\bm{0})$. In the Higgs rest frame the photon angular dependence is isotropic, 
and the integration over $d\Omega_1$ is trivial. Dropping the prefactors, the differential scattering rate becomes
\begin{equation}	
	d\Gamma \propto |\mathcal{M}|^2 \bigg[\mathop{\prod_{i = 1,2}}_{\alpha = \pm} |\bm{p}_{i_\alpha}| d\Omega_{i_\alpha}\bigg]dE_{1_-}dE_{2_-}~.\label{dGamma_BH}
\end{equation}
We may now proceed to consider CPV observables.

\subsection{Global sensitivity to CP violation}
\label{sec:CPVS}

In principle all the information about $\ctilde\ne 0$ (or equivalently $\xi\ne 0$) is encoded in the full HBH differential distribution. The coefficient $\ctilde$ may be determined by a matrix element method \cite{Abazov:2004cs,Abazov20051,Artoisenet:2010cn,Dalitz:1992dp,Fiedler:2010sg,JPSJ.60.836}, as long as backgrounds can be kept under control. Estimating the full power of the matrix element method is beyond the scope of this work. 

\begin{figure}\centering
	\includegraphics[scale=0.75]{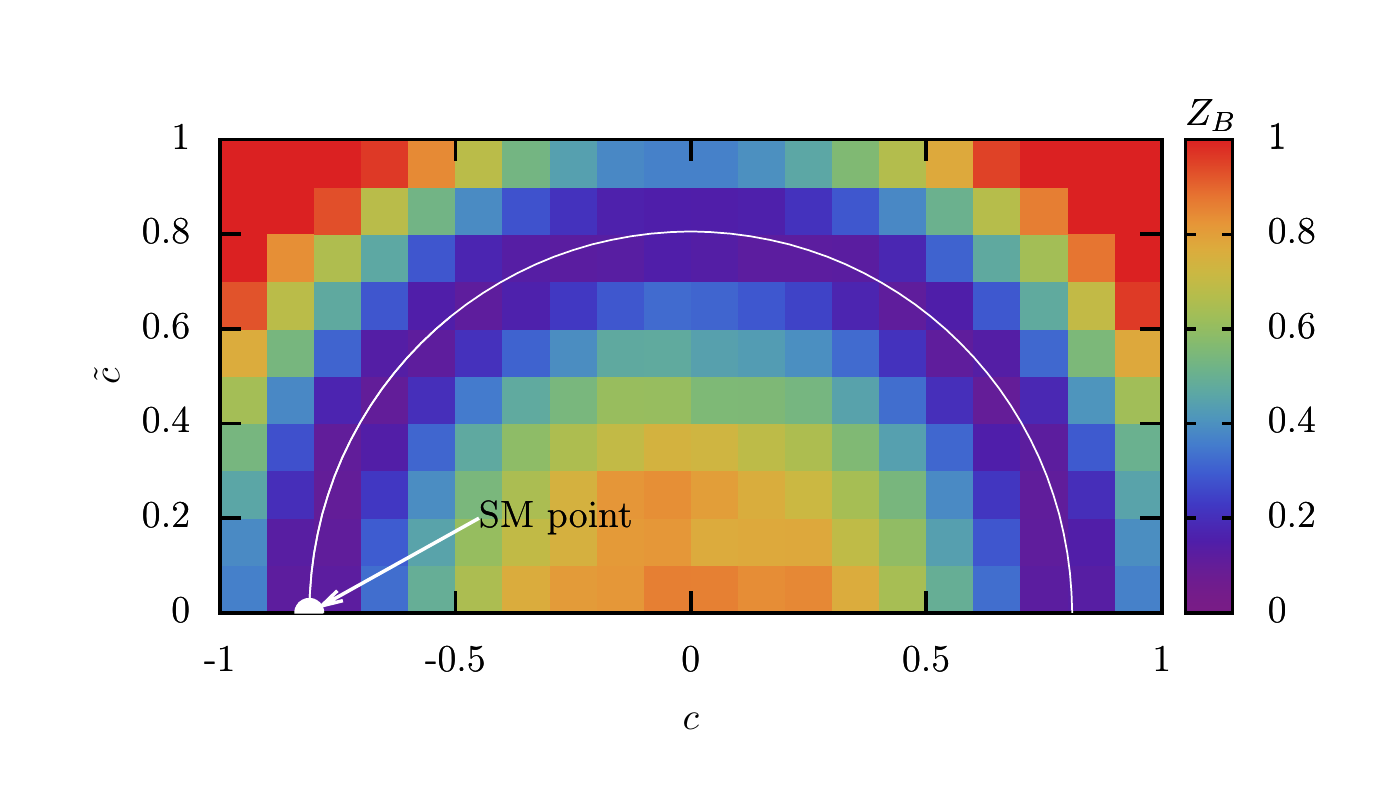}
	\includegraphics[scale=0.75]{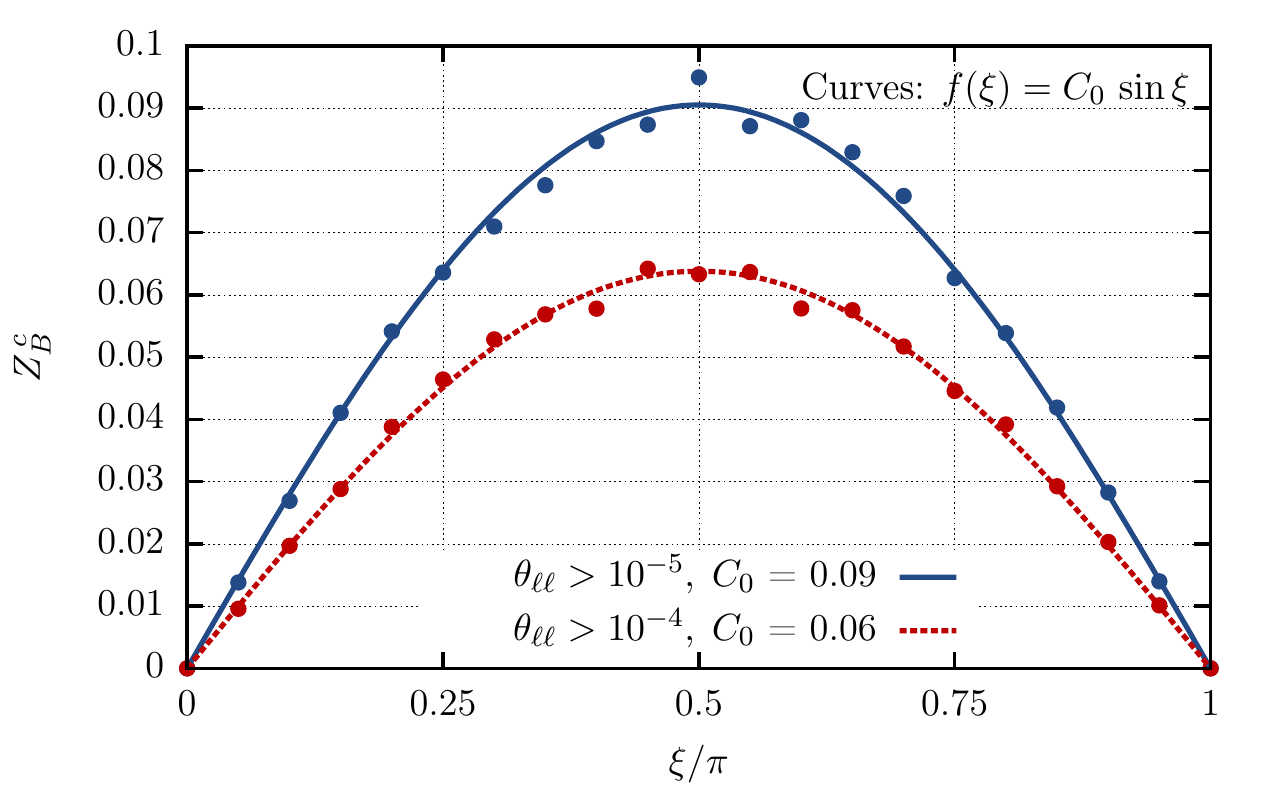}
\caption{ Top panel: $Z_B$ in the $c-\tilde c$ plane, with the SM point at $(c,\tilde c)=(-0.81,0)$. Bottom panel: $Z_B^c$ as a function of $\xi=\tan^{-1}(\tilde c/c)$. The scatter of the data points is a numerical artifact.
}
\label{fig:ZB-matrix}
\end{figure}

To test the sensitivity of HBH to $\tilde c$ we instead introduce a parameter
\begin{align}
	Z_B (c , \ctilde ) & = \frac{\int \left|d\Gamma(\text{\sc sm})/\dPS - d\Gamma(c, \tilde c)/\dPS \right|\dPS}{\Gamma(\text{\sc sm})}~.
\end{align}
The parameter $Z_B$ can be thought of as a proxy for the sensitivity of the matrix element method, once one integrates over the full phase space. In the top panel of Fig. \ref{fig:ZB-matrix} we show the value of $Z_B$ in the $(c, \tilde c)$ plane. There we see that the deviation from the SM is mostly due to the $c^2+\tilde c^2$ enhancement of the $h\to \gamma\gamma$ rate, which need not arise from CP violation. Such an enhancement is best detected by measuring the overall $\hgg$ rate, and not using our method.

To assess the sensitivity to CP violation alone, we should restrict our attention to the circular contour $c^2+\tilde c^2=c^2_{\mathrm{SM}}$, on which the total HBH rate matches the SM rate for any $\xi$. This contour is shown as a white circle in the top panel of Fig.~\ref{fig:ZB-matrix}. To this end, we define a second quantity
\begin{equation}
	\label{eqn:DZA}
	Z_B^{\rm c} (\xi)  \equiv Z_B(\xi)\bigg|_{c^2+\tilde c^2=c^2_{\mathrm{SM}}} = \frac{\int \left|{d\hat\Gamma}(0)/\dPS - {d\hat \Gamma}(\xi)/\dPS \right|\dPS}{{\hat\Gamma}(0)}~,
\end{equation}
where $d\hat\Gamma(\xi)$ is the differential HBH rate with the enhancement of the total $h\to \gamma\gamma$ rate factored out,
\begin{equation}
	d\Gamma(c,\tilde c)/\dPS = \frac{(c^2 + \ctilde^2)}{c_{\rm SM}^2}{d\hat\Gamma}(\xi)/\dPS.
\end{equation}
Note that in general $Z_B\in [0,\infty)$, but $Z_B^{\rm c}\in [0,2]$.

The bottom panel in Fig. \ref{fig:ZB-matrix} shows the value of $Z_B^{\rm c}$ as a function of $\xi$. The sinusoidal dependence of $Z_B^{\rm c}$ on $\xi$  is not unexpected. For instance, in the $\hgg$  toy example we considered in Sec. \ref{sec:ATE}, in which the angle between linear polarizations is measured, one has from Eq.~\eqref{eqn:HGGDR}
\begin{equation}
	Z_B^{\rm c}(\xi)\bigg|_{\hgg} = \frac{1}{\pi} \int_0^{2\pi} | \cos^2(\phi + \xi) - \cos^2(\phi)| d\phi = \frac{4}{\pi}|\sin \xi\,|~.
\end{equation}
By comparing this toy system with HBH, it is therefore natural to deduce that $Z_B^{\rm c}(\xi)/ |\sin\xi|$ provides a measure of the average CPV signal size obtainable via the matrix element method, for any $c, \ctilde$. That is, this measure is independent of $\xi$ and the overall normalization $c^2 + \ctilde^2$. The fit in Fig. \ref{fig:ZB-matrix} suggests this CPV signal is $\mathcal{O}(10\%)$ for an angular resolution cut $\theta_{\ell\ell} > 10^{-5}$, and $\mathcal{O}(5\%)$ for $\theta_{\ell\ell} > 10^{-4}$. Hence, for the most pessimistic case that there is no deviation from the SM in the total $h\to \gamma\gamma$ rate the detection of NP from the full matrix element method will be challenging, even if there is large CPV component in $\hgg$.

\subsection{Differential azimuthal scattering rate}
Let us also consider the sensitivity of an experiment in which just one relative azimuthal angle -- the difference of the azimuthal angles between two opposite branch leptons -- is reconstructed, such as the experiment described in Fig. \ref{fig:cartoon}.  From Eqs.~\eqref{eqn:HBHA} and \eqref{eqn:BHSALOA} we saw that $\xi$ manifests as a phase shift in the relative azimuthal orientation of leptons on different branches. It is instructive to write down this manifestation explicitly. 
Let us transform from the azimuthal coordinates $\phi_{i_\pm}$ to the coordinates 
\begin{equation}
	\label{eqn:DVP}
	\varphi \equiv \phi_{1_+} + \phi_{2_+}~,\qquad \varepsilon_{i} \equiv \phi_{i_-} - \phi_{i_+}~,
\end{equation}
and choose $\phi_{1_+} = 0$, without loss of generality. With this choice $\varphi$ and $\varepsilon_1 \in [0,2\pi)$, while $\varepsilon_2 \in (-2\pi,2\pi)$. For the case that the Higgs is at rest, we find from Eqs.~\eqref{eqn:HBHAE} and \eqref{eqn:BHSALOA} that the HBH square amplitude, and thus the leading order differential scattering rate, takes the generic form
\begin{equation}
	\label{eqn:GPSGT}
	d\Gamma/\dPS = \frac{1}{q_1^4q_2^4}\bigg[ a + \sum_j b_j \cos\varepsilon_j + \sum_k c_k \cos\big(n_k \varepsilon_1 + m_k \varepsilon_2 + 2\varphi + 2\xi\big)\bigg]~,
\end{equation}
where $a$, $b_j$ and $c_{k}$ are real functions of $\gamma_{i_\pm}$ and $\theta_{i_\pm}$ -- they span the energy, polar angle phase space, denoted hereafter by $\textrm{\sc PS}_{\gamma,\theta}$ --  but are independent of the azimuthal structure, and $n_k$, $m_k$ are positive integers that satisfy $n_k + m_k = 0, 1$, or $2$. From Eq.~\eqref{eqn:QPAE}, one sees that the $1/q^4$ factors depend on both the azimuthal and polar angles, but in a way such that $q^4 = \eta(1 + \zeta\cos\varepsilon)^2$, where $\zeta < 1$. Hence $1/q^4$ may be expanded in a power series of $\zeta \cos \varepsilon < 1$. Integrating over the $\varepsilon$ acoplanarity angles, one may then show that, with respect to the azimuthal structure, only constant or $\cos(2\varphi + 2\xi)$ terms survive. That is, the leading order differential scattering rate
\begin{equation}
	\label{eqn:GVPD}
	 \frac{d\Gamma}{d\varphi~ \dPS_{\gamma,\theta}} = (c^2 + \ctilde^2)\Big[ \mathcal{A}_{\gamma,\theta} + \mathcal{B}_{\gamma,\theta}\cos(2\varphi + 2 \xi)\Big]~,
\end{equation}
in which the $\gamma,\theta$ subscript denotes exclusive dependence on the energy polar angle phase space, $\textrm{\sc PS}_{\gamma,\theta}$. 

The results \eqref{eqn:GPSGT} or \eqref{eqn:GVPD} show that $\xi$-dependence and $\gamma,\theta$-dependence factorize. Hence $\xi$ appears only in the azimuthal structure as a phase shift. Specifically, the CPV parameter $\xi$ manifests itself in the inter-branch azimuthal structure in the differential rate
\begin{equation}
	\frac{d\Gamma}{d\varphi} = (c^2 + \ctilde^2)\Big[ \big\langle  \mathcal{A}_{\gamma,\theta} \big\rangle_{\textrm{\sc PS}_{\gamma,\theta}} + \big\langle  \mathcal{B}_{\gamma,\theta} \big\rangle_{\textrm{\sc PS}_{\gamma,\theta}}\cos(2\varphi + 2 \xi)\Big]~,
\end{equation}
which we have now shown is oscillatory with respect to $\varphi$ at leading order. Note that we could have chosen $\varphi$ to be any of the four inter-branch angles $\varphi_{1_\alpha 2_\beta} \equiv \phi_{1_\alpha} + \phi_{2_\beta}$, where $\alpha$, $\beta = \pm$, of which three are linearly independent. Results similar to Eq.~\eqref{eqn:GVPD} follow with appropriate replacements.

\begin{figure}[t]\centering
	\includegraphics[width=0.45\columnwidth]{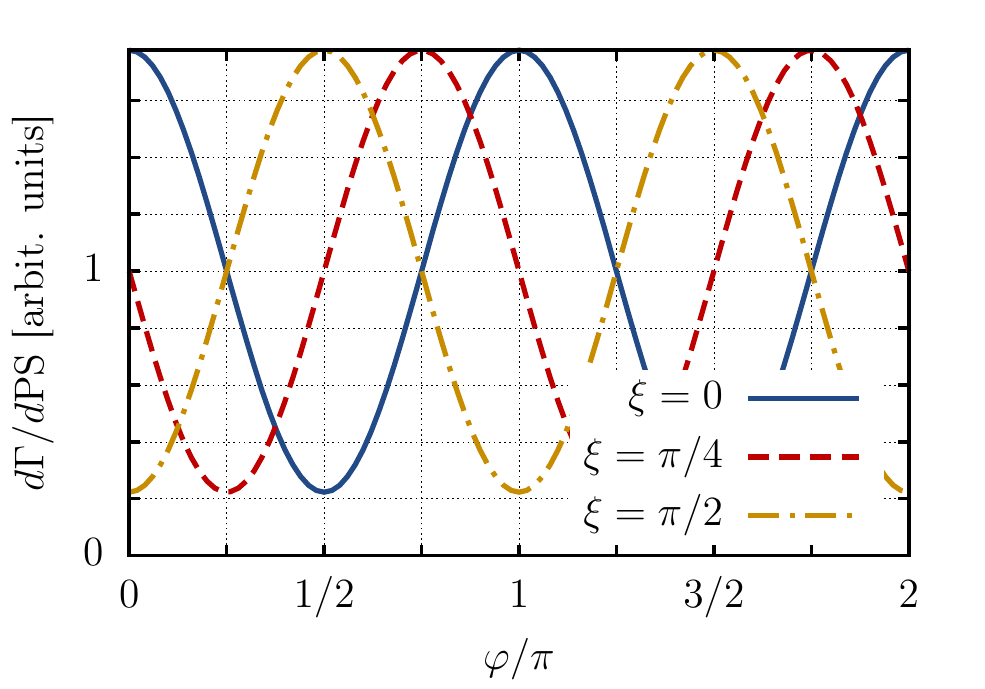}
	\includegraphics[width=0.45\columnwidth]{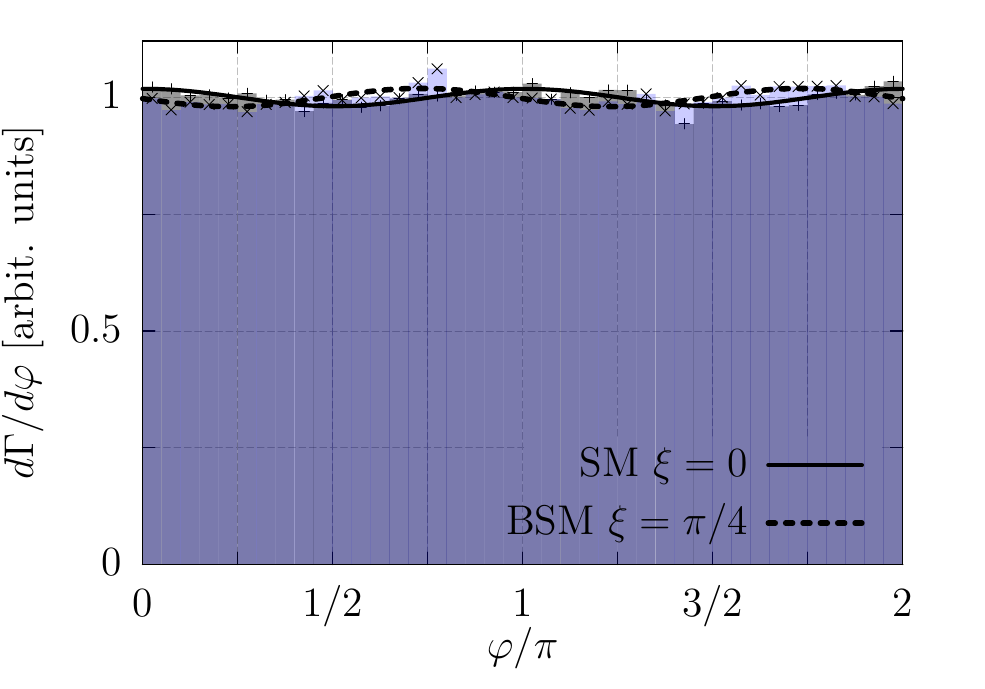}
\caption{Left: Illustration of $\mathcal{O}(1)$ oscillations and phase shifts in the HBH differential rate for a sample coplanar kinematic configuration. The azimuthal angle $\varphi$ in this slice is defined as in Eq.~\eqref{eqn:DVP}. The kinematic configuration is: $E_{i_+}=E_{i_-}=m_h/4,\;\theta_{i_+}=10^{-4},\;\theta_{i_-}=2\theta_{i_+}$ so that $\gamma_{\pm}\theta_{\pm} \sim 10 \gg 1$ and $\gamma_+ = \gamma_-$, cf. analysis of Eq.~\eqref{eqn:ERBA}. Right: The azimuthal distribution $d\Gamma/d\varphi$ for $\xi=0$ and for $\xi = \pi/4$ with a polar angle cut $\theta_{i_\pm} > 10^{-5}$ and $\theta_{\ell\ell} > 10^{-5}$. The modulation amplitude is $2\%$, but will grow to $\mathcal{O}(1)$ once optimization cuts are applied, see Sec.~\ref{sec:EC}.  
}
\label{fig:PSslice-azim}
\end{figure}

In the right panel of Fig.~\ref{fig:PSslice-azim} we show the differential distribution $d\Gamma/d\varphi$ for HBH events (including a loose cut on the polar angles, at $10^{-5}$) for two values of $\xi$. It is evident that the oscillation amplitude, $\BAv{}$, is small -- approximately $\sim 2\%$ -- when averaged over all of the phase space $\textrm{\sc PS}_{\gamma,\theta}$.

The key question we now wish to address is whether such a small oscillation amplitude is because of small, $\mathcal{O}(1\%)$ oscillations, or whether there is an $\mathcal{O}(1\%)$ part of the phase space where deviations from the SM are ${\mathcal O}(1)$.  In the language of Eq.~\eqref{eqn:GVPD}, this question can be rephrased in a precise manner: Is $\BAv{U}$ small for all $U \subset \textrm{\sc PS}$, or does there exist $U \subset \textrm{\sc PS}$ such that $\BAv{U}$ is $\mathcal{O}(1)$. The latter possibility is phenomenologically preferred, since it permits the extraction of an $\mathcal{O}(1)$ CPV signal on $U$, which would scale better with increasing statistics.

To address this question, let us begin by examining the coplanar limit. In this limit the acoplanarity angles $\varepsilon_{1,2}$ are both zero, and we have from Eq.~\eqref{eqn:GPSGT}
\begin{equation}
	\label{eqn:CDSA}
	\frac{d\Gamma}{d\varphi~ \dPS_{\gamma,\theta}}  = \mathcal{A}^\mathrm{co}_{\gamma,\theta} + \mathcal{B}^\mathrm{co}_{\gamma,\theta}\cos(2\xi + 2\varphi)~. 
\end{equation}
In the coplanar limit the size of the modulation is given by
\begin{equation}
	\label{eqn:ERBA}
	\frac{\mathcal{B}^\mathrm{co}_{\gamma,\theta}}{\mathcal{A}^\mathrm{co}_{\gamma,\theta}} = \prod_{i = 1,2}  \frac{ \mathcal{R}_i(1 - \gamma_{i_+}\theta_{i_+} \gamma_{i_-}\theta_{i_-})^2}{(1 +\gamma^2_{i_+}\theta^2_{i_+} )(1 + \gamma^2_{i_-}\theta^2_{i_-} ) + \mathcal{R}_i(\gamma_{i_-}\theta_{i_-} +\gamma_{i_+}\theta_{i_+} )^2}~,
\end{equation}
where $\mathcal{R}_i \equiv 2 \gamma_{i_+}\gamma_{i_-}/(\gamma_{i_+}^2 + \gamma_{i_-}^2)$. The ratio $\mathcal{B}^\mathrm{co}/\mathcal{A}^\mathrm{co}$ is small when $\gamma\theta \sim 1$, i.e. near the peak of the square matrix element, but $\mathcal{B}^\mathrm{co}/\mathcal{A}^\mathrm{co} \to 1$, for $\gamma\theta \gg 1$ and $\gamma_\pm$ not much bigger than $\gamma_\mp$. An example is shown in Fig. \ref{fig:PSslice-azim}, where  $E_{i_\pm}$ and $\theta_{i_\pm}$ are held fixed such that $\theta_{i_\pm} \gg m/E_{i_\pm}$ and $\gamma_+ \sim \gamma_-$. In this slice of phase space the azimuthal oscillation amplitude is large and there is a strong sensitivity to CPV. This shows that regions of phase space with large CPV signals exist.

\subsection{ CPV enhancing cuts} 
\label{sec:EC}

We now use the results from Sec.~\ref{sec:hel} to design kinematic cuts that enhance the sensitivity of $d\Gamma/d\varphi$ to CP violation. That is, we seek to enhance the ratio $\BAv{}$ for a particular subset of the HBH event sample. The cuts we propose fall into two classes: those which are placed on the kinematics of the whole event; and those which are placed independently on individual photons and their daughter leptons. As they use all of the information in the event, including correlations among the two photons, one might expect that the former produce better CPV signals compared to the latter. As we shall see, however, both classes of cuts perform approximately equally well in enhancing $\BAv{}$ on their respective phase space subregions.

\subsubsection{Cuts on collective kinematics}

The helicity interference terms (\ref{eqn:CPOE}) in the HBH rate arise dominantly from the term $\mbox{Re}( \alpha_{1_{12}}^-\alpha_{1_{21}}^-\alpha_{2_{12}}^-\alpha_{2_{21}}^-e^{-2i\xi})$. We can use this observation to pick only events in which $\alpha_{1_{12}}^-\alpha_{1_{21}}^-\alpha_{2_{12}}^-\alpha_{2_{21}}^-$ is comparable to the rest of the squared amplitude. With reference to Eqs.~\eqref{eqn:CPO} and \eqref{eqn:CPE} we thus introduce several sensitivity parameters $\mathcal{T}_n$
\begin{equation}
	\mathcal{T}_n  \equiv \mathcal{X}_n \Bigg/\Bigg[\Big|\alpha^-_{1_{11}}\alpha^-_{2_{11}}\Big|^2   +  \mathop{\mathop{\sum}_{j \not= k}}_{r_{k}\not=s_{k}}\!\!\Big|\alpha^-_{j_{11}}\alpha^-_{k_{rs}}\Big|^2 +  \mathcal{X}_n\Bigg],
	\label{eqn:TPD}
\end{equation}
where $n = M^{(\prime)},R^{(\prime)},J^{(\prime)}$, with
\begin{gather}
	\mathcal{X}_\mathrm{M}  = 4|\alpha^-_{1_{12}}\alpha^-_{1_{21}}\alpha^-_{2_{12}}\alpha^-_{2_{21}}|~, \quad 
	\mathcal{X}_\mathrm{R}  = 4\mbox{Re}[ \alpha^-_{1_{12}}\alpha^-_{1_{21}}\alpha^-_{2_{12}}\alpha^-_{2_{21}}]~, \quad 
	\mathcal{X}_\mathrm{J}  = 4\mbox{Im}[ \alpha^-_{1_{12}}\alpha^-_{1_{21}}\alpha^-_{2_{12}}\alpha^-_{2_{21}}]~,\notag\\
	\mbox{and} \qquad \mathcal{X}_\mathrm{M',R',J'} = \mathop{\mathop{\sum}_{j \not= k}}_{r\not=s,\rho\not=\sigma}\!\!\Big|\alpha^-_{j_{rs}}\alpha^-_{k_{\rho\sigma}}\Big|^2 + \mathcal{X}_\mathrm{M,R,J}~.
\end{gather}
The first two terms of the denominator in Eq.~\eqref{eqn:TPD} are simply the $\xi$ independent parts of the HBH squared amplitude (\ref{eqn:HBHAE}), while the $\mathcal{X}_n$'s are various pieces of the interference terms. In particular, $\mathcal{T}_\mathrm{M}$ is the magnitude of the full leading order interference term, while $\mathcal{T}_\mathrm{R,J}$ are respectively the CP-even and CP-odd interference terms. We expect each to be useful gauge of sensitivity to CPV independent of $\xi$. For example, cuts on $\mathcal{T}_\mathrm{R}$ and $\mathcal{T}_\mathrm{J}$ will enhance the azimuthal modulations in the $\xi=0$ and $\xi=\pi/4$ respectively, and can thus be used to distinguish among these. In all cases $\mathcal{T}_i \to 1$ ($\mathcal{T}_i \to 0$) implies full (no) CPV sensitivity.

\begin{figure}
\centering
\includegraphics[width=0.49\columnwidth]{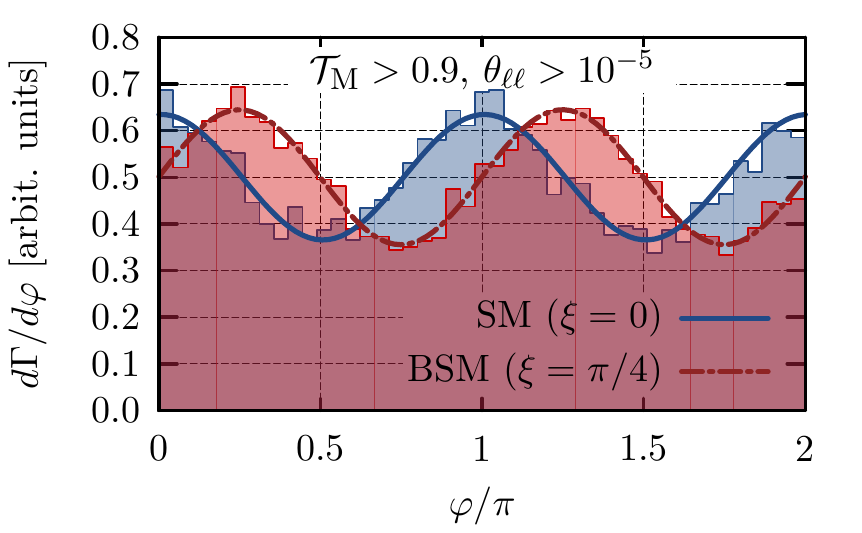}
\centering
\includegraphics[width=0.49\columnwidth]{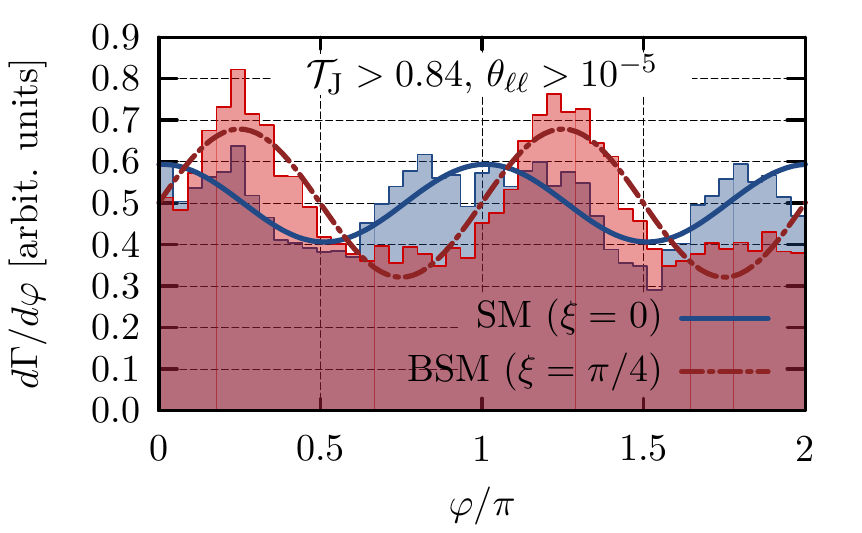}
\centering
\includegraphics[width=0.49\columnwidth]{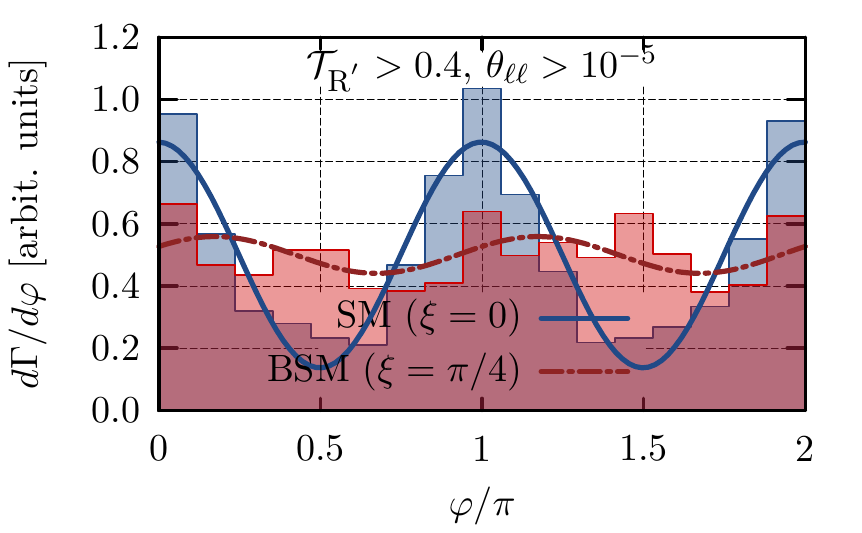}
\centering
\includegraphics[width=0.49\columnwidth]{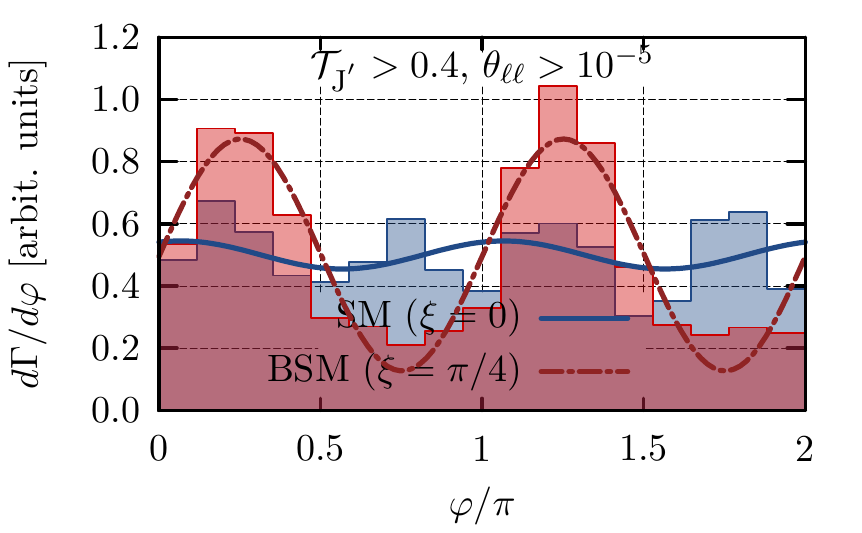}
\caption{The differential HBH rate $d\Gamma/d\varphi$ as a function of the azimuthal angle $\varphi$ between the positrons setting the CPV parameter to $\xi=0$ (blue histograms) and $\xi=\pi/4$ (red histograms). The histograms are binned Monte Carlo events with different cuts on parameters $\mathcal{T}_n$, Eq.~\eqref{eqn:TPD}, as denoted in the panels. 
The solid blue (red dot-dashed) curves are the result of fitting the normalized binned events to Eq.~\eqref{eqn:CDSA} for the $\xi=0$ ($\xi=\pi/4$) cases with $\xi$ also floated in the fit. The angular resolution cut is $\theta_{\ell\ell} > \theta_{\rm cut}=10^{-5}$.}
\label{fig:DiffAzim10-5}
\end{figure}

\begin{figure}
\centering
\includegraphics[width=0.49\columnwidth]{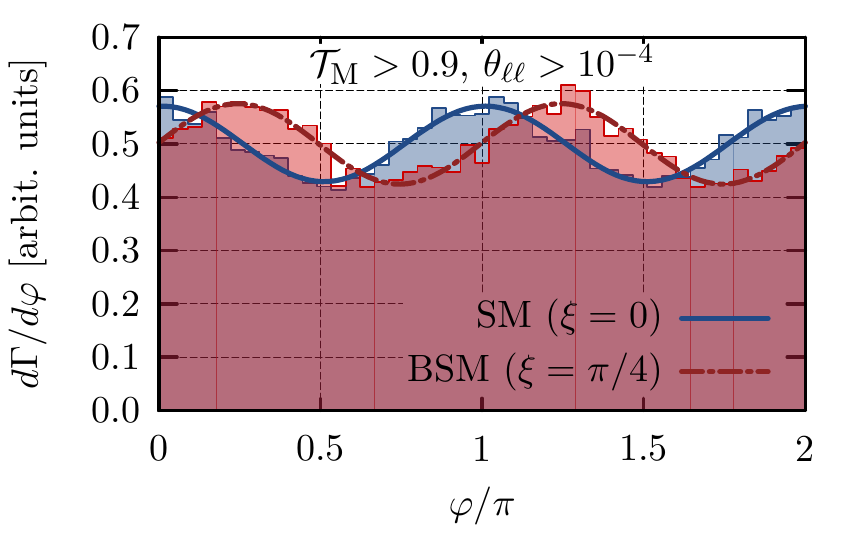}
\centering
\includegraphics[width=0.49\columnwidth]{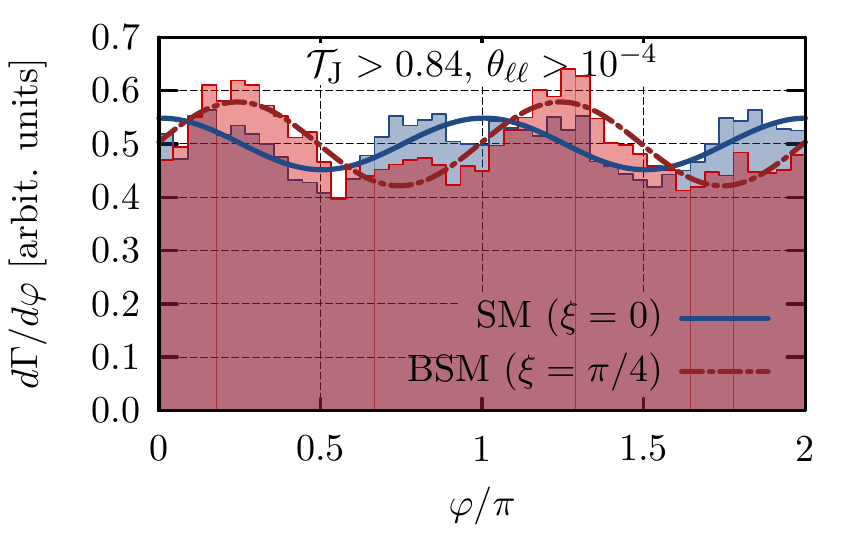}
\\
\centering
\includegraphics[width=0.49\columnwidth]{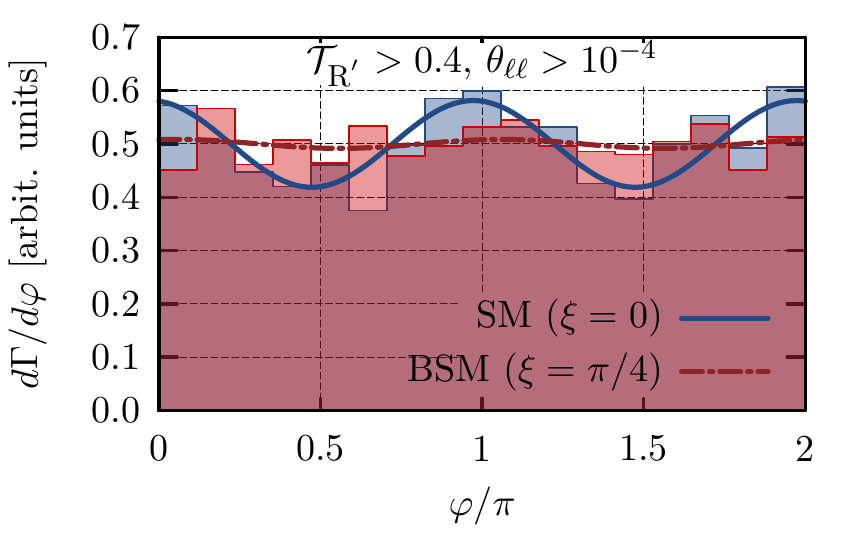}
\centering
\includegraphics[width=0.49\columnwidth]{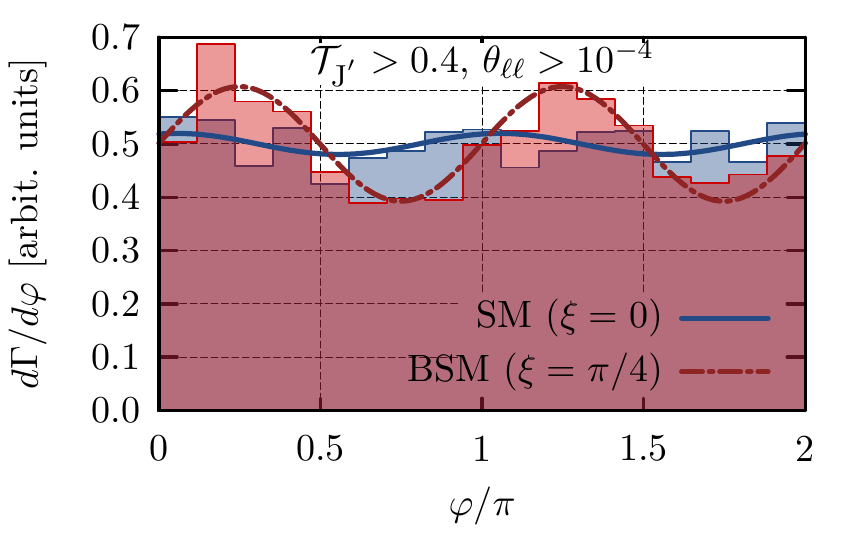}
\caption{The same as in Fig. \ref{fig:DiffAzim10-5}, but with the opening angle cut $\theta_{\ell\ell} > \theta_{\rm cut}=10^{-4}$.}
\label{fig:DiffAzimVar}
\end{figure}

For all numerical analysis we use three private Monte Carlo codes that were cross-checked. The details on Monte Carlo event generation can be found in Appendix \ref{app:MCgen}. Placing a cut on $\mathcal{T}_n$ produces an event sample with large $\BAv{}$ ratio. This is shown in Fig.~\ref{fig:DiffAzim10-5} for opening angle cut $\theta_{\ell\ell}>10^{-5}$ and in Fig.~\ref{fig:DiffAzimVar} for opening angle cut of $\theta_{\ell\ell}>10^{-4}$. The $d\Gamma/d\varphi$ distributions are shown for two choices of CPV parameters, $\xi=0$ (blue histograms) and $\xi=\pi/4$ (red histograms). 
The fits to the functional form \eqref{eqn:GVPD} of the $d\Gamma/d\varphi$ HBH differential rate are shown as solid blue and dot-dashed red lines, respectively. The efficiencies of the cuts for the examples shown in Figs.~\ref{fig:DiffAzim10-5} and~\ref{fig:DiffAzimVar} are $\sim1\%$ for the upper panels and $\sim 0.1\%$ for the lower panels (the exact values of efficiencies depend on the value of $\xi$).  The presence of $\xi\ne0$ exhibits itself as the expected phase shift in $d\Gamma/d\varphi$ differential rate \eqref{eqn:GVPD}.


From Figs.~\ref{fig:DiffAzim10-5} and \ref{fig:DiffAzimVar} we see that it is possible to select regions of phase space such that $\BAv{} \sim {\mathcal O}(1)$. In particular the average modulation $\BAv{}$ is large (small) for $\xi=0$ ($\xi=\pi/4$) for ${\mathcal T}_\mathrm{R,R'}$ and vice versa for ${\mathcal T}_\mathrm{J,J'}$ as shown most strikingly in the bottom panels. In contrast, the average modulation size does not depend on $\xi$ for the events selected by ${\mathcal T}_\mathrm{M}$. This suggests that using several of the parameters $\mathcal{T}_n$ simultaneously may optimize the sensitivity to CPV further.

\subsubsection{ Cuts on individual photon conversions}

We now turn to discuss cuts on individual photon branches of the HBH process. These cuts are generated by simple sufficiency conditions, that ensure a large CPV signal. 

From Eqs.~\eqref{eqn:HBHAE} and \eqref{eqn:CPOE}, in order to ensure that the $\xi$-dependent terms are comparable to the full rate, it suffices to require $|\alpha^-_{i_{12}}\alpha^-_{i_{21}}| \gtrsim |\alpha^-_{i_{11}}|^2$ on each branch $(i = 1,2)$ independently. For the case that the Higgs is at rest, using \eqref{eqn:BHSALOA}, this sufficiency condition is implied by $\bS \gtrsim 1$ for each branch, where
\begin{equation}
	\label{eqn:SBSC}
	 \bS \equiv 2(1-\cos\delta)\bigg[\frac{\gamma_+\gamma_-}{(\gamma_+ + \gamma_-)^2}\bigg]
	\bigg[
	\frac{ \gamma_+ \theta_+ \gamma_-\theta_- }{(\gamma_+^2\theta_+^2 - \gamma_-^2\theta_-^2)^2} 
	\bigg] (1 + \gamma_+^2\theta_+^2)(1+ \gamma_-^2\theta_-^2)~.
\end{equation}
For brevity, we have dropped the Latin branch index, and here $\delta \equiv |\phi_- -\phi_+| -\pi$ is the acoplanarity angle. 

It is notable that this sufficiency condition weights the desirable energy-polar angle regions of phase space inversely by the amount of acoplanarity. Specifically, $\bS = 0$ in the exact coplanar configuration $\delta = 0$. Of course, $\bS$ is not a necessary condition for a large CPV signal, so relatively coplanar events -- i.e. events for which $\delta \ll \delta_{\rm cut}$ of the event sample ($\delta_{\rm cut} = 0.4\pi$ or $0.25\pi$ for our MC, see App. \ref{app:MCgen}) -- may in principle significantly contribute to large CPV signals in $d\Gamma/d\varphi$. Also note that symmetric conversions, i.e. $\gamma_+\theta_+ \simeq \gamma_-\theta_-$, are more likely to produce a large $\bS$, and hence a strong interference effect.

\begin{figure}[t]\centering
\includegraphics[width=0.49\columnwidth]{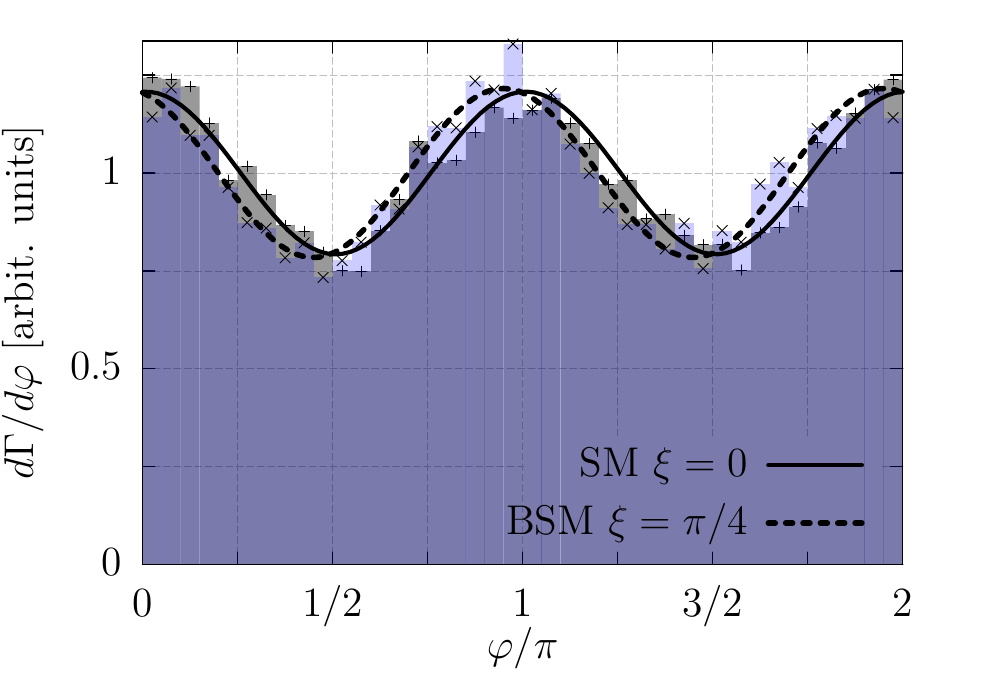}
\includegraphics[width=0.49\columnwidth]{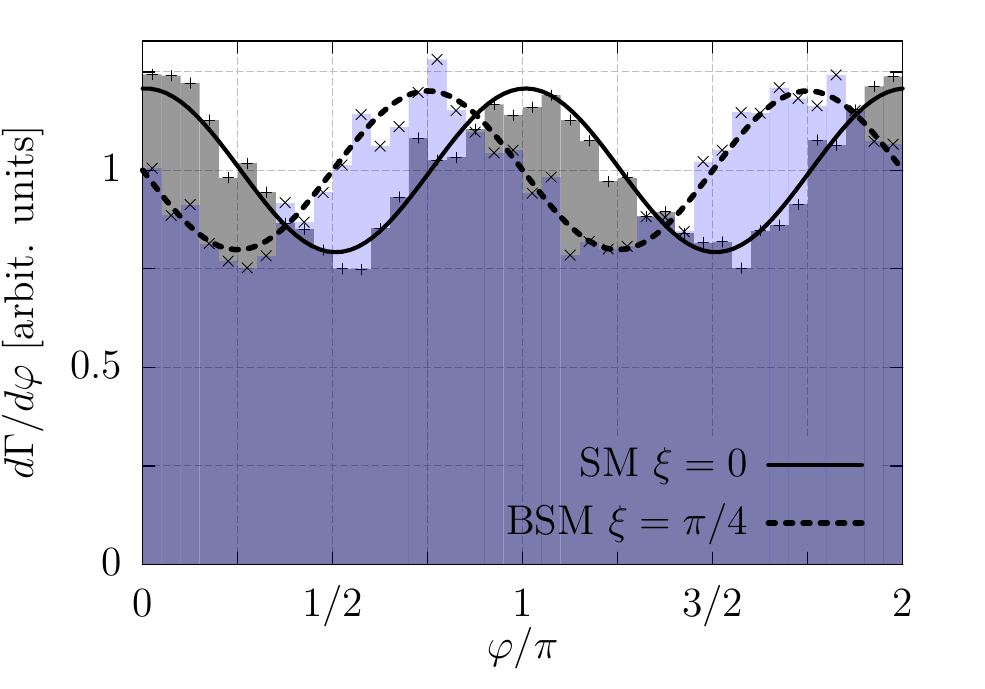}
\\
\centering
\includegraphics[width=0.49\columnwidth]{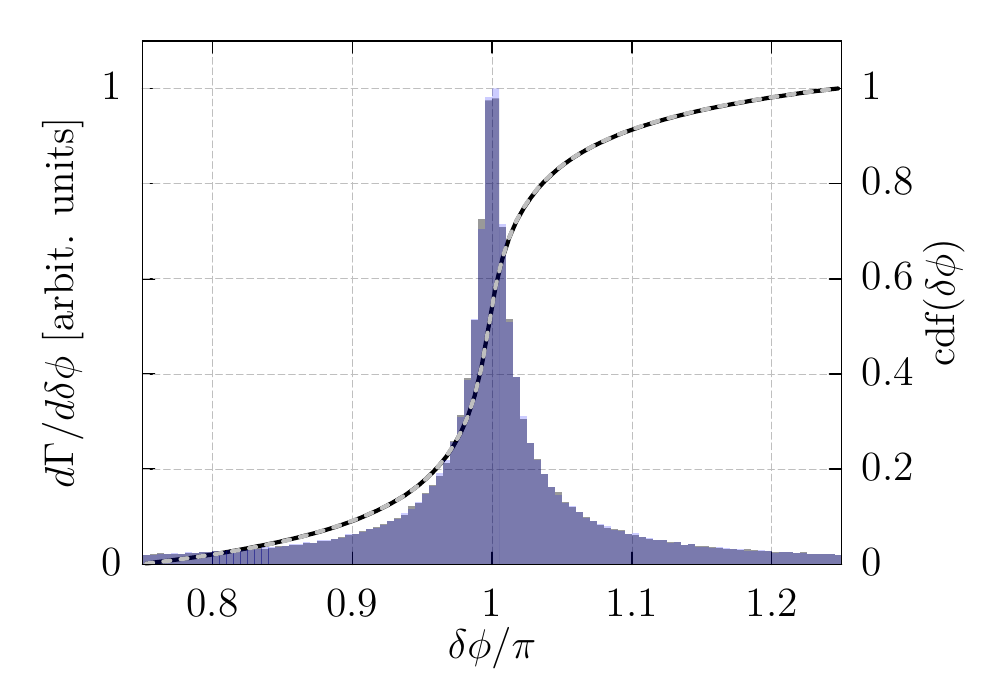}
\includegraphics[width=0.49\columnwidth]{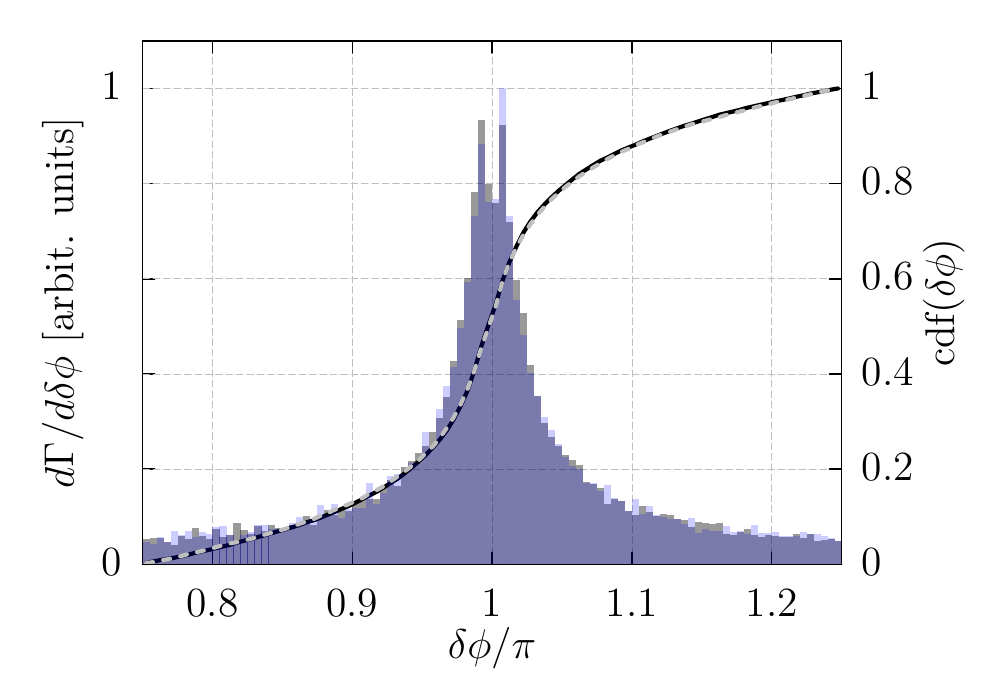}
\caption{Top panels: The azimuthal distributions $d\Gamma/d\varphi$ for $\xi=0$ (grey histograms) and for $\xi = \pi/20, \pi/4$  (blue histograms, left and right top panels) with $\bS_{1,2} >1$ on the domain $\delta_{1,2} \in [3\pi/4,5\pi/4]$. The solid (dashed) curves denote fits to Eq.~\eqref{eqn:GVPD}, with $\xi$ a free parameter in the fits. The bottom left (right) panel shows the acoplanarity distributions, $d\Gamma/d\delta\phi$, $\delta\phi \equiv (\phi_+ - \phi_-) \mod 2\pi$ for each photon branch, displayed by blue and gray histograms respectively, with $\xi = \pi/4$ and no $\bS$ cut ($\bS_{1,2} > 1$). Note that the scale varies between these two plots. The corresponding cumulative acoplanarity distributions  $\mbox{cdf}(\delta\phi)$ for each branch are denoted by solid black and grey dashed lines on each plot. In all panels the angular resolution cut \eqref{eqn:ARC} of $\theta_{\rm cut} = 10^{-5}$ was applied. 
}
\label{fig:SBC}
\end{figure}

Fig. \ref{fig:SBC} shows the results of the $\bS_{1,2}>1$ cut for the angular resolution cut \eqref{eqn:ARC} $\theta_{\rm cut} = 10^{-5}$ for $\xi = 0$ compared to $\xi = \pi/20$ (top left) or $\pi/4$ (top right). Note we also constrain the acoplanarity $\delta_{1,2} \in [3\pi/4, 5\pi/4]$, since we do not expect extremely acoplanar events to encode polarization information. The angular resolution cut alone retains approximately $40\%$ of the total HBH events on this acoplanar domain; for $\xi = \pi/4$, the corresponding azimuthal distribution is shown in Fig. \ref{fig:PSslice-azim}. The addition of the $\bS_{1,2} > 1$ cut reduces the cut efficiency to $2.6\%$, but unlike Fig. \ref{fig:PSslice-azim}, now $ \BAv{} \sim {\mathcal O}(0.2)$. The phase shift due to non-zero $\xi$ is clearly visible in both plots, and the value of $\xi$ extracted from the fits agrees with the input values of $\xi = \pi/20$ or $\pi/4$ respectively. 

Finally, the bottom panels of Fig. \ref{fig:SBC} displays the acoplanarity on each branch for the $\xi = \pi/4$ HBH events with no $\bS$ cut compared to the $\bS_{1,2} >1$ cuts. We see that the $\bS$ cut mildly favors acoplanar events, as it broadens the acoplanarity distribution and disfavors relatively coplanar events. For example, we see that events with acoplanarity $\delta < 1\%$ are disfavored in the cut distribution, and we also see that the full width at half maximum of the acoplanar distribution increases by $50\%$, from $0.04\pi$ to $0.06\pi$, under the $\bS> 1$ cut. The excellent performance of $\bS$ compared to $\mathcal{T}_n$ cuts (see Fig. \ref{fig:CSC} and Sec. \ref{sec:CSC} below) therefore suggests that acoplanar events -- e.g. with $\delta \not\ll \delta_{\rm cut}$ -- play an important role in encoding the CPV signal.

\subsubsection{Cut scheme efficiencies}
\label{sec:CSC}
\begin{figure}[t]\centering
\includegraphics[width =0.49\columnwidth]{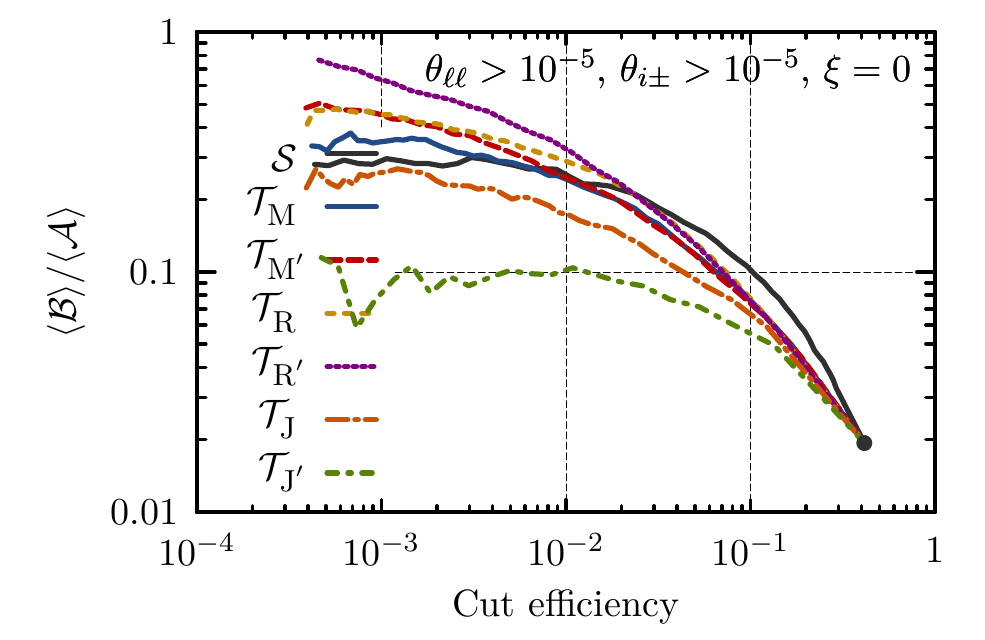}
\includegraphics[width =0.49\columnwidth]{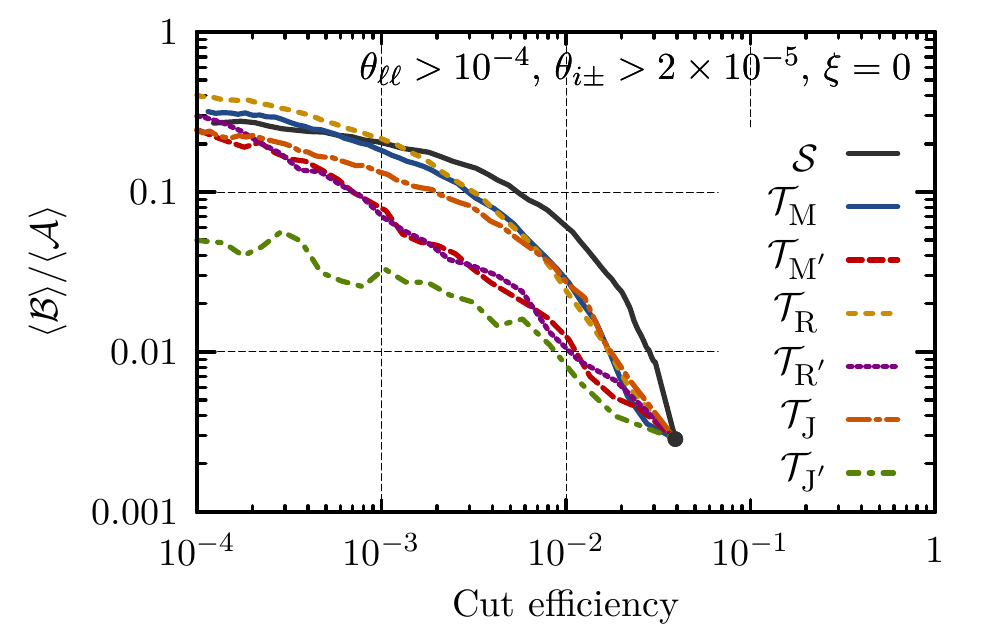}
\\
\includegraphics[width =0.49\columnwidth]{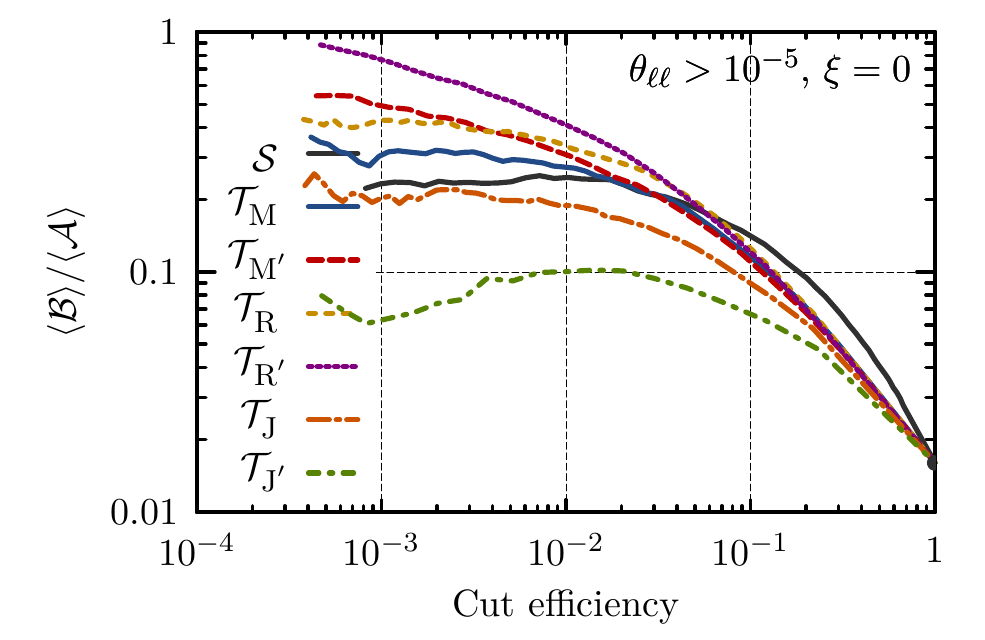}
\includegraphics[width =0.49\columnwidth]{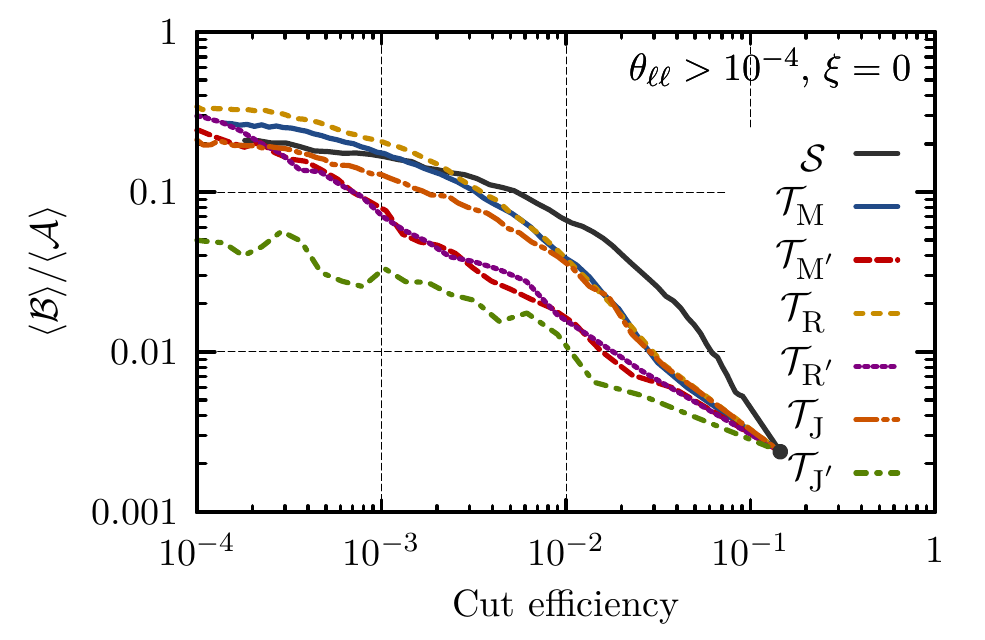}
\caption{Comparison of different sensitivity parameter cut schemes, in terms of cut efficiency, for various angular resolution cuts. The black data point on each plot denotes the efficiency and $\BAv{}$ for its angular resolution cut alone, with no enhancements from sensitivity parameter cuts.}
\label{fig:CSC}
\end{figure}

It remains to determine the efficiency of the above cut schemes. In Fig. \ref{fig:CSC} we show the CPV signal $\BAv{}$ for the various $\mathcal{T}_n$ and $\bS$ schemes as a function of the fraction of the total MC event sample, not rejected by combined sensitivity and angular resolution cuts. Up to small corrections this fraction is the absolute cut efficiency, see Appendix \ref{app:MCgen} for details. For comparison, the data point with only angular resolution cuts is also shown for each plot.

We see that an increase of the CPV signal by an order of magnitude roughly requires an order of magnitude penalty in sample size. Moreover, for high cut efficiencies, the $\bS_i$ scheme provides the largest CPV signal. In contrast, in the low efficiency, but higher signal regions, the $\mathcal{T}_n$ schemes outperform the $\bS_i$ cuts. To give an idea about the feasibility of this analysis at the (HL/HE-)LHC, Table~\ref{tab:numevts} gives the expected number of events after the application of the cuts discussed above. These numbers do not inlcude experimental acceptance efficiencies and so will be lower in practice.

The $\bS_i$ parameter is defined on an individual photon branch, independently of the lepton configurations on the other branch, while $\mathcal{T}_n$ is defined on the configuration of whole HBH events. We therefore might deduce that the main difference between these two schemes is whether they are affected by inter-branch leptonic configurations. If this deduction is correct, then the comparative performance of $\bS_i$ versus $\mathcal{T}_n$ cuts, shown in Fig. \ref{fig:CSC}, suggest that inter-branch configurations become more important for the extraction of larger CPV signals, but are not important over most of the CPV sensitive phase space. 

\begin{table}[b!]\centering
\begin{tabular}{|c|c|c|c|}\hline
$\sqrt{s}$ & $\mathcal{L}$ [fb$^{-1}$] & $\sigma\times\text{BR}(h\rightarrow\gamma\gamma)$ [fb] & Events\\\hline\hline
8 & 20 & 47 & 0.24\\\hline
14 & 3000 & 125 & 94\\\hline
33 & 3000 & 444 & 333\\\hline
100 & 3000 & 1875 & 1406\\\hline
\end{tabular}
\caption{Expected number of events after the application of $\mathcal{S}$ or $\mathcal{T}$ cuts with $\theta_{\ell\ell}>10^{-4}$ to obtain $\langle \mathcal{B}\rangle/\langle\mathcal{A}\rangle\sim 20\%$. The Higgs production cross section includes the gluon fusion and VBF channels only and is taken from~\cite{hxswg}.}
\label{tab:numevts}
\end{table}

\subsection{Triple products}
Finally, we briefly comment on the possibility to use CP-odd quantities such as triple products to directly search for CPV.  Since the CP odd terms in $|\mathcal{M}|^2$ are $C$ even and $P$ odd, we could consider two such contractions
\begin{equation}
	\mathcal{\tau}_1 = \epsilon^{\Pm1 \Pp1 \Pm2 \Pp2},
\end{equation}
and
\begin{equation}
	 \mathcal{\tau}_2 = \frac{1}{4}\big(\epsilon^{P \Pp1 \Pm2 \Pp2} + \epsilon^{ \Pm1P \Pm2 \Pp2} + \epsilon^{\Pm1 \Pp1 P \Pp2} + \epsilon^{\Pm1 \Pp1 \Pm2 P}\big)~,
\end{equation}
where $P^\mu = (M,\bm{0})$ is the nucleus momentum, and $\epsilon^{pqrs}$ is shorthand for the Levi-Civita contraction $\epsilon^{\mu\nu\rho\sigma}p_\mu q_\nu r_\rho s_\sigma$.
In terms of scalar triple products, 
\beq
	\begin{split}
	\mathcal{\tau}_1  &= E_{1_+}\bm{p}_{1_-}\!\cdot\!(\bm{p}_{2_+}\!\times\bm{p}_{2_-}) - E_{1_-}\bm{p}_{1_+}\!\cdot\!(\bm{p}_{2_+}\!\times\bm{p}_{2_-}) +1\leftrightarrow 2~,\\
	\mathcal{\tau}_2  &= M\Big[\bm{p}_{1_-}\cdot(\bm{p}_{2_+}\times\bm{p}_{2_-}) - \bm{p}_{1_+}\cdot(\bm{p}_{2_+}\times\bm{p}_{2_-}) +  1\leftrightarrow 2\Big]~.
	\end{split}
\eeq
 Note, that $\tau_2$ is not strictly speaking C-even, since it involves the nucleus momentum. However, BH conversion on nucleus is the same as on anti-nucleus within our uncertainties. This is equivalent to leaving $P^\mu$ unchanged under C transformation.  A detailed analysis is beyond the scope of our work, but we remark in passing that a straightforward use of the $\bS$ and ${\mathcal T}_n$ cuts does not lead to appreciable non-zero value of $\langle \tau\rangle$. Further investigation is warranted.

\section{Conclusion}
In this work we have studied how to probe the underlying CP couplings of the Higgs to two photons, which undergo Bethe-Heitler conversion on nuclei. We have shown that sensitivity to CP violating couplings is possible only if there is interference between conversion amplitudes with different photon helicity. Using spinor helicity methods, we have computed compact, leading order expressions for these amplitudes, which are novel to this work. Our analytical control of the leading order $\hgg \stackrel{BH}{\to} 4 e$ full differential scattering rate permits us to show that: (i) the differential rate with respect to the relative azimuthal angles between leptons with different parent photons is oscillatory; (ii) CPV is encoded as a phase shift in such distributions; and (iii) we may construct various sensitivity parameter cuts that extract the regions of phase space on which such oscillations -- the CPV signal -- are order unity on average. These analytical results have been confirmed and explored with numerical simulations, including a comparison of the relative effectiveness of the different sensitivity parameters.

For simplicity we restricted our numerical and CP sensitivity analysis to the case that the Higgs is at rest in the lab frame. This is not the case in the LHC experiments, and this assumption needs to be lifted for more realistic studies.  We note, in this vein, that Eqs.~\eqref{eqn:BHSALO} hold also for the boosted Higgs case, although Eqs.~\eqref{eqn:BHSALOA} do not. The sensitivity parameter cut schemes are expected to work comparatively well in the case that the Higgs is boosted, too. On a similar note, we expect these sensitivity cuts to enhance the full matrix element method. This method, as characterized by the parameter $Z_B^{\rm c}$, appears to be a few times more sensitive to CPV than experiments measuring a single relative azimuthal angle. This leaves open the possibility of further improved CPV signals, compared to those shown in this work.

We do not expect a $\hgg \stackrel{BH}{\to} 4 e$ experiment to be straightforward, and further experimental and theoretical studies are needed in order to see if the methods discussed in this paper can be used in practice. From an experimental viewpoint, it needs to be determined how well one can reconstruct the electron and positron momenta for opening angles $\sim 10^{-4}$ or even down to $\sim m/m_h \sim 10^{-5}$, and whether there are significant rescatterings after BH production. Already from our preliminary studies it is clear that large statistical datasets and fine granularity of the detectors will be needed, such as at the proposed HE-LHC, VLHC or TLEP~\cite{Dawson:2013bba}. Experimentally, the situation may be more favorable in such a machine, having a larger amount of statistics and better kinematic control of the Higgs. A completely independent direction for measuring the CP violating coupling of Higgs to photons -- and a direction not explored in this paper -- would be the use of polarized photon beams at a photon collider. From a theoretical viewpoint, it also remains to determine and search for other CPV sensitive observables apart from the azimuthal distributions discussed here.

Lastly, while we focussed on Higgs diphoton decays in this work, most of this analysis holds for any pair of correlated photons. They can arise from a resonance, like the
Higgs, or from scattering events. While the details have to be determined in a case-by-case basis, the principles, such as helicity interference, are the same.

\acknowledgments{
We thank Yang Bai, Kfir Blum, John-Paul Chou, Lance Dixon, Andrei Gritsan, Ben Heidenreich, Kirill Melnikov, Michael Peskin, Mike Sokoloff, Roberto Vega-Morales and Ciaran Williams for helpful discussions. We also expressly thank Jacob Meisler for the development of the Monte-Carlo unweighting  \texttt{Java/C++} code, used in part of the numerical analysis in this work. FB is supported in part by the Fermilab Fellowship in Theoretical Physics. The work of YG and DR is supported by the U.S. National Science Foundation through grant PHY-0757868 and by the United States-Israel Binational Science Foundation (BSF) under grant No.~2010221. The work of DR is also supported by the U.S. National Science Foundation under Grant No.~PHY-1002399. JZ was supported in part by the U.S. National Science Foundation under CAREER Grant PHY-1151392.  We thank the Aspen Center for Physics and the NSF Grant \#1066293 for hospitality during the conception of this work. We would also like to thank KITP for warm hospitality during the completion of this work and acknowledge that
this research was supported in part by the National Science Foundation under Grant No. NSF PHY11-25915. Fermilab is operated by the Fermi Research
Alliance, LLC under Contract No.~De-AC02-07CH11359 with the United States Department of Energy.
}

\appendix

\section{Spinor-helicity formalism}
\label{app:SHF}

In this paper we extensively use the spinor-helicity formalism, in which the sigma matrices $\sigma^\mu_{a\dot{a}}$ solder null momenta to Weyl spinors, that transform under the spinor irreducible representations of the covering group SL(2,$\mathbb{C}$). For a review see e.g. Ref. \cite{Dixon:1996wi} or \cite{Dixon:2004za}.

For a null momentum $k$, the associated Weyl spinors $\lambda_k$ are soldered via
\begin{equation}
	k^\mu = \frac{1}{2}\sigbr{\pm}{k}{\sigma^\mu}{k}~, \qquad \slashed{k} \equiv k_\mu \sigma^\mu_{a\dot{a}} = (\lambda_k)_a(\lambda_k)_{\dot{a}}~,
\end{equation}
in which we have written the dotted and undotted Weyl spinors in the notation
\begin{equation}
	\langle k^-| \equiv (\lambda_k)^a~, \qquad \langle k^+| \equiv (\lambda_k)_{\dot{a}}~, \qquad |k^+\rangle \equiv (\lambda_k)_{a}~, \qquad |k^-\rangle \equiv (\lambda_k)^{\dot{a}}~,
\end{equation}
so $\sigbr{+}{k}{\sigma^\mu}{\ell} = (\lambda_k)_{\dot{a}} \sigma^{\mu \dot{a}a}(\lambda_\ell)_a$,  $\sigbr{-}{k}{\sigma^\mu}{\ell} = (\lambda_k)^{a} \sigma^{\mu}_{a\dot{a}}(\lambda_\ell)^{\dot{a}}$. Upper and lower indices are related by the usual epsilon contractions, in particular $\sigma_\mu^{\dot{a}a} \equiv \epsilon^{ab}\epsilon^{\dot{a}\dot{b}}{\sigma_\mu}_{b\dot{b}}$. In this notation, the sign superscripts denote the helicities of the in-going and out-going states, corresponding to kets and bras respectively. In particular, note that $| k^\pm \rangle^\dagger = \langle k^\pm |$ by construction, $\langle k^\pm |\sigma^\mu |\ell^\mp \rangle$ do not exist, while $\sigbr{-}{k}{\sigma^\mu}{\ell} = \sigbr{+}{\ell}{\sigma^\mu}{k}$. For null momenta $k$ and $\ell$ we often write the bilinears
\begin{equation}
	\agbr{ k}{ \ell } \equiv \langle k^- | \ell^+\rangle = (\lambda_{k})^a(\lambda_{\ell})_a~,\qquad \sqbr{ k }{ \ell } \equiv \langle k^+ | \ell^-\rangle = (\lambda_{k})_{\dot{a}}(\lambda_{\ell})^{\dot{a}}~,
\end{equation}
so $\agbr{\ell}{k} = -\agbr{k}{\ell}$, $\sqbr{\ell}{k} = -\sqbr{k}{\ell}$, $\agbr{ k }{ \ell } \sqbr{ \ell }{ k }  = 2 k \cdot \ell$ and $\agbr{ k}{ \ell } ^* = \sqbr{ \ell }{ k } $. 

In this notation, definite helicity polarization vectors associated to $k$ may have form
\begin{equation}
	\label{eqn:SHPV}
	\varepsilon^{\pm\mu}(k; r) = \pm\frac{\langle r^\mp | \sigma^\mu | k ^\mp\rangle}{\sqrt{2} \langle r^\mp | k^\pm \rangle}~,
\end{equation}
with $r^\mu$ a reference null momentum. One may show via the Fierz relations that these polarization vectors satisfy the axiomatic requirements $k\cdot \varepsilon^\pm =0$, $\varepsilon^\pm \cdot (\varepsilon^\pm)^* = \varepsilon^\pm \cdot \varepsilon^\mp = -1$ and  $\varepsilon^\pm \cdot \varepsilon^\pm = 0$. In the present work, a particularly convenient choice of reference momenta is 
\begin{equation}
	\label{eqn:RMC}	
	r_1 = k_2~, \qquad r_2 = k_1~, 
\end{equation}
where $r_{1,2}$ ($k_{1,2}$) are reference momenta (photon momenta) corresponding to polarization vectors $\varepsilon^\pm_{1,2}$ defined in \eqref{eqn:HGGHA}. We shall always make this choice of reference momenta.

The dotted and undotted Weyl spinors have explicit representations in terms of the momentum components in a particular basis, up to a free choice of phase. For example, for a null momentum $k = (k^0, k^1,k^2,k^3)$, a possible choice is
\begin{equation}
	\label{eqn:WSPC}
	(\lambda_k)^a = \begin{cases} \begin{pmatrix} \dfrac{k^1 + i k^2}{\sqrt{k^0 - k^3}} ~,& \sqrt{k^0 - k^3} \end{pmatrix}~, & k^3 < 0~, \\[7pt] \begin{pmatrix} \sqrt{k^0 + k^3} ~, & \dfrac{k^1 - i k^2}{\sqrt{k^0 + k^3}} \end{pmatrix}~, & k^3 >0 ~.\end{cases}
\end{equation}
The components of the upper dotted spinor are immediately specified through the relation $(\lambda_k)^{\dot{a}} = [(\lambda_k)^{a}]^\dagger$, and the lower index spinors through appropriate epsilon contractions. With this choice and a choice of basis, all spinor objects -- e.g. $\agbr{k}{\ell}$ -- can be evaluated explicitly, just as momentum objects may be, e.g. $k\cdot \ell$. Note that for our Weyl spinor phase choice \eqref{eqn:WSPC} one has $\agbr{k_1}{k_2} = \sqbr{k_2}{k_1} = m_h$.

The Dirac spinors of massive particles may also be represented in the spinor-helicity formalism, via the application of a light-cone decomposition. For a massive spinor of momentum $p$ and mass $m$, we define an associated null momentum
\begin{equation}
	\tilde{p}^\mu = p^\mu - \frac{m^2}{2 p \cdot \ell} \ell^\mu~, \quad\mbox{or equivalently} \quad p^\mu = \tilde{p}^\mu + \frac{m^2}{2 \tilde{p} \cdot \ell} \ell^\mu~,
\end{equation}
where $\ell^\mu$ is a null reference momentum. The spinors then decompose similarly as
\begin{align}
	u^1(p) & = |\tilde{p}^+\rangle + \frac{m}{\sqbr{\tilde{p} }{\ell} } |\ell^-\rangle, & u^2(p) & = |\tilde{p}^-\rangle - \frac{m}{\agbr{ \ell }{ \tilde{p}} } |\ell^+\rangle,\notag\\
	\bar{u}^1(p) & = \langle \tilde{p}^+| + \frac{m}{\agbr{ \ell  }{\tilde{p}}} \langle \ell^-|, & \bar{u}^2(p) & = \langle \tilde{p}^-| - \frac{m}{\sqbr{\tilde{p} }{\ell } } \langle \ell^+|,\notag\\
	v^1(p) & = |\tilde{p}^-\rangle + \frac{m}{\agbr{\ell }{\tilde{p} }} |\ell^+\rangle, & v^2(p) & = |\tilde{p}^+\rangle - \frac{m}{\sqbr{\tilde{p} }{\ell } } |\ell^-\rangle,\notag\\
	\bar{v}^1(p) & = \langle \tilde{p}^-| + \frac{m}{\sqbr{ \tilde{p} }{\ell}} \langle \ell^+|, & \bar{v}^2(p) & = \langle \tilde{p}^+| - \frac{m}{\agbr{\ell}{ \tilde{p} } } \langle \ell^-|~. \label{eqn:SLCD}
\end{align}
One may verify that these spinors satisfy the canonical requirements $\bar{u}u = 2m$, $\bar{v}v = -2m$, $\bar{v}u = \bar{u}v = 0$ and the completeness relations $\sum_ju^j\bar{u}^j = \slashed{p} + m$ and  $\sum_jv^j\bar{v}^j = \slashed{p} - m$. Just as for the polarization vectors, one is free to choose the null reference momentum. This choice amounts to a choice of `gauge', under which the unpolarized square amplitude must be invariant, but the polarized square amplitudes are not. In this work, we shall always make the reference momenta choices
\begin{equation}
	\label{eqn:WSRM}
	\ell_{1_\pm} = k_2~ \quad \mbox{and} \quad \ell_{2_\pm} = k_1~.
\end{equation}	
These are convenient choices for the purposes of extracting the leading order BH helicity amplitudes, as below.

Finally, together with the several well-known spinor identities, the following identity is especially useful for computing terms involving Levi-Civita contractions,
\begin{equation}	
	\label{eqn:ESI}
	\epsilon^{\alpha\beta\gamma\delta}\sigma_\alpha^{\dot{a}a}\sigma_\beta^{\dot{b}b}\sigma_\gamma^{\dot{c}c}\sigma_\delta^{\dot{d}d} \equiv 4i\big( \epsilon^{\dot{d}\dot{a}}\epsilon^{ba}\epsilon^{\dot{b}\dot{c}}\epsilon^{dc} - \epsilon^{\dot{d}\dot{c}}\epsilon^{bc}\epsilon^{\dot{b}\dot{a}}\epsilon^{da}\big)~.
\end{equation}

\section{BH spin-helicity amplitudes}
\label{app:SHA}
Here we write down explicit expressions for the helicity amplitudes  $\alpha^\pm_{i_\rs}$. From Eqs.~\eqref{eqn:FFSA} and \eqref{eqn:MQA} one has (suppressing the branch index)
\begin{align}
	\alpha_{\rs}^\pm
	& = -\frac{i}{2q^2}\bar{u}^r(p_-)\bigg[ \slashed{\varepsilon}^{\pm}\frac{\slashed{p}_{-} - \slashed{k} + m}{k \cdot p_{-}} \slashed{Q} + \slashed{Q}\frac{\slashed{k}- \slashed{p}_{+}  + m}{k\cdot p_{+}}\slashed{\varepsilon}^{\pm} \bigg] v^s(p_{+})\notag\\
	& = -\frac{i}{2q^2}\bigg[ \sum_j \bigg(\frac{\bar{u}^r\slashed{\varepsilon}^{\pm} u^j\bar{u}^j \slashed{Q} v^s}{k\cdot p_{-}}  - \frac{\bar{u}^r\slashed{Q} v^j\bar{v}^j \slashed{\varepsilon}^{\pm} v^s}{k\cdot p_{+}}\bigg)
	- \frac{\bar{u}^r\slashed{\varepsilon}^{\pm} \slashed{k}\slashed{Q}v^s}{k\cdot p_{-}} +  \frac{\bar{u}^r \slashed{Q} \slashed{k} \slashed{\varepsilon}^{\pm} v^s}{k\cdot p_{+}}\bigg]~,
\end{align}
where $u = u(p_-)$ and $v = v(p_+)$. Applying the light cone decomposition (\ref{eqn:SLCD}) with reference choices specified in (\ref{eqn:SHPV}) and \eqref{eqn:WSRM}, gives the full results for each spin helicity amplitude
\begin{align}
\alpha^-_{11}  = - (\alpha^+_{22})^* 
& = \frac{1}{q^2\sqrt{2}}\bigg[-\frac{m \left[\ell\, \tilde{p}_{-}\right] \left\langle k\, \tilde{p}_{-}\right\rangle 
   \left\langle \ell{}^-\right|\slashed{Q}|\tilde{p}_{+}{}^-\rangle }{ k\cdot p_{-}
   \left[k\, \ell\right] \left\langle \ell\, \tilde{p}_{-}\right\rangle }+\frac{m \left[\ell\,
   \tilde{p}_{+}\right] \left\langle k\, \tilde{p}_{+}\right\rangle  \left\langle
   \ell{}^-\right|\slashed{Q}|\tilde{p}_{+}{}^-\rangle }{ k\cdot p_{+} \left[k\,
   \ell\right] \left\langle \ell\, \tilde{p}_{-}\right\rangle }\notag\\
   & \quad -\frac{m \left[\ell\,
   \tilde{p}_{-}\right] \left\langle k\, \tilde{p}_{-}\right\rangle  \left\langle
   \tilde{p}_{-}{}^+\right|\slashed{Q}|\ell{}^+\rangle }{ k\cdot p_{-} \left[k\,
   \ell\right] \left\langle \ell\, \tilde{p}_{+}\right\rangle } +\frac{m \left[\ell\,
   \tilde{p}_{+}\right] \left\langle k\, \tilde{p}_{+}\right\rangle  \left\langle
   \tilde{p}_{-}{}^+\right|\slashed{Q}|\ell{}^+\rangle }{ k\cdot p_{+} \left[k\,
   \ell\right] \left\langle \ell\, \tilde{p}_{+}\right\rangle }\notag\\
   &\quad -\frac{m \left\langle k\,
   \ell\right\rangle  \left\langle k{}^-\right|\slashed{Q}|\tilde{p}_{+}{}^-\rangle }{
   k\cdot p_{-} \left\langle \ell\, \tilde{p}_{-}\right\rangle }+\frac{m \left\langle k\,
   \ell\right\rangle  \left\langle \tilde{p}_{-}{}^+\right|\slashed{Q}|k{}^+\rangle }{
   k\cdot p_{+} \left\langle \ell\, \tilde{p}_{+}\right\rangle }\bigg]\notag\\
\alpha^-_{12}  = + (\alpha^+_{21})^* 
& = \frac{1}{q^2\sqrt{2}}\bigg[ -\frac{m^2 \left\langle k\, \ell\right\rangle  \left\langle k{}^-\right|\slashed{Q}|\ell{}^-\rangle }{ k\cdot p_{-} \left[\ell\, \tilde{p}_{+}\right] \left\langle
   \ell\, \tilde{p}_{-}\right\rangle }
   +\frac{m^2 \left\langle k\, \tilde{p}_{+}\right\rangle 
   \left\langle \ell{}^-\right|\slashed{Q}|\ell{}^-\rangle }{ k\cdot p_{+} \left[k\,
   \ell\right] \left\langle \ell\, \tilde{p}_{-}\right\rangle }\notag\\
   & \quad -\frac{m^2 \left[\ell\,
   \tilde{p}_{-}\right] \left\langle k\, \tilde{p}_{-}\right\rangle  \left\langle
   \ell{}^-\right|\slashed{Q}|\ell{}^-\rangle }{ k\cdot p_{-} \left[k\, \ell\right]
   \left[\ell\, \tilde{p}_{+}\right] \left\langle \ell\, \tilde{p}_{-}\right\rangle
   }
   -\frac{\left[\ell\, \tilde{p}_{-}\right] \left\langle k\, \tilde{p}_{-}\right\rangle 
   \left\langle \tilde{p}_{-}{}^+\right|\slashed{Q}|\tilde{p}_{+}{}^+\rangle }{
   k\cdot p_{-} \left[k\, \ell\right]}\notag\\
  & \quad   +\frac{\left[\ell\, \tilde{p}_{+}\right] \left\langle
   k\, \tilde{p}_{+}\right\rangle  \left\langle \tilde{p}_{-}{}^+\right|\slashed{Q}|\tilde{p}_{+}{}^+\rangle }{ k\cdot p_{+} \left[k\,
   \ell\right]}
   +\frac{\left\langle k\, \tilde{p}_{+}\right\rangle  \left\langle
   \tilde{p}_{-}{}^+\right|\slashed{Q}|k{}^+\rangle }{ k\cdot p_{+}}\bigg]\notag\\
\alpha^-_{21}  = + (\alpha^+_{12})^* 
& = \frac{1}{q^2\sqrt{2}}\bigg[ -\frac{\left[k\, \tilde{p}_{-}\right] \left\langle k\, \tilde{p}_{-}\right\rangle 
   \left\langle \tilde{p}_{-}{}^-\right|\slashed{Q}|\tilde{p}_{+}{}^-\rangle }{
   k \cdot p_{-} \left[k \, \ell \right]}
   +\frac{\left[\ell \, \tilde{p}_{+}\right] \left\langle
   k\, \tilde{p}_{+}\right\rangle  \left\langle \tilde{p}_{-}{}^-\right|\slashed{Q}|\tilde{p}_{+}{}^-\rangle }{ k\cdot p_{+} \left[k \, \ell \right]}\notag\\
  & \quad  +\frac{m^2
   \left\langle k\, \ell\right\rangle  \left\langle \ell{}^+\right|\slashed{Q}|k{}^+\rangle
   }{ k\cdot p_{+} \left[\ell\, \tilde{p}_{-}\right] \left\langle \ell\,
   \tilde{p}_{+}\right\rangle }
   +\frac{m^2 \left[\ell\, \tilde{p}_{+}\right] \left\langle k\,
   \tilde{p}_{+}\right\rangle  \left\langle \ell{}^+\right|\slashed{Q}|\ell{}^+\rangle }{
   k\cdot p_{+} \left[k\, \ell\right] \left[\ell\, \tilde{p}_{-}\right] \left\langle \ell\,
   \tilde{p}_{+}\right\rangle }\notag\\
   & \quad -\frac{m^2 \left\langle k\, \tilde{p}_{-}\right\rangle 
   \left\langle \ell{}^+\right|\slashed{Q}|\ell{}^+\rangle }{ k\cdot p_{-} \left[k\,
   \ell\right] \left\langle \ell\, \tilde{p}_{+}\right\rangle }
   -\frac{\left\langle k\,
   \tilde{p}_{-}\right\rangle  \left\langle k{}^-\right|\slashed{Q}|\tilde{p}_{+}{}^-\rangle
   }{ k\cdot p_{-}}\bigg]\notag\\
\alpha^-_{22}  = -(\alpha^+_{11})^*    
& = \frac{1}{q^2\sqrt{2}}\bigg[  -\frac{m \left\langle k\, \tilde{p}_{-}\right\rangle  \left\langle k{}^-\right|\slashed{Q}|\ell{}^-\rangle }{ k\cdot p_{-} \left[\ell \, \tilde{p}_{+}\right]}
+\frac{m
   \left\langle k \, \tilde{p}_{+}\right\rangle  \left\langle \tilde{p}_{-}{}^-\right|\slashed{Q}|\ell{}^-\rangle }{ k \cdot p_{+} \left[k \, \ell \right]}\notag\\
   & \quad -\frac{m \left[\ell \,
   \tilde{p}_{-}\right] \left\langle k\, \tilde{p}_{-}\right\rangle  \left\langle
   \tilde{p}_{-}{}^-\right|\slashed{Q}|\ell {}^-\rangle }{ k \cdot p_{-} \left[k\,
   \ell \right] \left[\ell \, \tilde{p}_{+}\right]}
   +\frac{m \left\langle k\,
   \tilde{p}_{+}\right\rangle  \left\langle \ell {}^+\right|\slashed{Q}|k{}^+\rangle }{
   k \cdot p_{+} \left[\ell \, \tilde{p}_{-}\right]}\notag\\
   &\quad +\frac{m \left[\ell \, \tilde{p}_{+}\right]
   \left\langle k \, \tilde{p}_{+}\right\rangle  \left\langle \ell{}^+\right|\slashed{Q}|\tilde{p}_{+}{}^+\rangle }{ k \cdot p_{+} \left[k \, \ell \right] \left[\ell \,
   \tilde{p}_{-}\right]}
   -\frac{m \left\langle k \, \tilde{p}_{-}\right\rangle  \left\langle
   \ell{}^+\right|\slashed{Q}|\tilde{p}_{+}{}^+\rangle }{ k \cdot p_{-} \left[k\,
   \ell \right]}\bigg]~.\label{eqn:BHSA}
\end{align}

As per the main text, we have dropped the photon subscripts, and $k,\ell$ = $k_1, k_2$ or $k_2,k_1$ for parent photon 1 and 2 respectively. Squaring these amplitudes, taking traces and summing, one obtains the full Bethe-Heitler square amplitude that is obtained by the usual Feynman methods. 

We can further extract dominant terms of the BH spin-helicity amplitudes by observing that if $q_i \ll m_h$, then $\agbr{k_1}{k_2} \gg \agbr{k_i}{p_{i_\pm}}$ etc. Moreover, in expressions such as $\alpha^+_{12}$ or $\alpha^+_{21}$, we may discard subdominant $\mathcal{O}(m^2/k_i\cdot p_{j\not=i})$  terms.   This leads to the following leading order results in $m^2/k_i\cdot p_{j\not=i}$ and $ \agbr{k_i}{p_{i}}/\agbr{k_1}{k_{2}}$ for the BH spin-helicity amplitudes
\begin{align}
	\alpha^-_{11}  = - (\alpha^+_{22})^* & \simeq \frac{m}{q^2\sqrt{2}}\bigg[\frac{\agbr{ k }{ \ell }  \sigbr{+}{ \tilde{p}_-} { \sigma^0 }{ k} }{k \cdot p_+ \agbr{\ell}{ \tilde{p}_+}} -\frac{\agbr{ k }{ \ell }  \sigbr{+}{ \tilde{p}_+}{  \sigma^0 }{ k} }{k \cdot p_- \agbr{\ell }{\tilde{p}_-}} \bigg],\notag\\
	\alpha^-_{12}  =  \phantom{-}(\alpha^+_{21})^* &  \simeq \frac{\agbr{ k }{\tilde{p}_{+}}  \sigbr{+} { \tilde{p}_{-}}{ \sigma^0}{k} }{q^2\sqrt{2} k \cdot p_{+}}
   -\frac{\sqbr{\ell }{ \tilde{p}_{-}} \agbr{ k}{ \tilde{p}_{-}}  \sigbr{+}{ \tilde{p}_{-}}{ \sigma^0}{\tilde{p}_{+}} }{q^2\sqrt{2} k \cdot p_{-} \sqbr{k }{\ell }}
   +\frac{ \sqbr{\ell }{ \tilde{p}_{+}} \agbr{ k}{ \tilde{p}_{+}}  \sigbr{+}{ \tilde{p}_{-}}{ \sigma^0}{\tilde{p}_{+}} }{q^2\sqrt{2} k \cdot p_{+} \sqbr{k}{ \ell }}\notag\\
   & \qquad \qquad -\frac{m^2 \agbr{ k  }{\ell } \sigbr{-}{ k}{\sigma^0}{\ell}}{q^2\sqrt{2} k \cdot p_{-} \sqbr{\ell}{ \tilde{p}_{+}} \agbr{ \ell }{\tilde{p}_{-}} },\notag\\
    \alpha^-_{21}  =   \phantom{-}(\alpha^+_{12})^* & \simeq  -  \frac{\agbr{ k  }{\tilde{p}_{-} } \sigbr{+}{ \tilde{p}_{+}}{ \sigma^0}{k}}{q^2\sqrt{2} k \cdot p_{-}} -\frac{\sqbr{\ell }{\tilde{p}_{-}} \agbr{ k }{\tilde{p}_{-}} \sigbr{-}{ \tilde{p}_-}{ \sigma^0 }{\tilde{p}_+}}{q^2\sqrt{2}  k\cdot p_{-} \sqbr{k }{\ell }}  + \frac{\sqbr{\ell }{\tilde{p}_+} \agbr{k }{\tilde{p}_+} \sigbr{-}{\tilde{p}_-}{\sigma^0 }{\tilde{p}_+}}{q^2\sqrt{2} k \cdot p_{+} \sqbr{ k }{\ell} } \notag\\
    & \qquad \qquad + \frac{m^2 \agbr{ k }{ \ell } \sigbr{-}{k}{\sigma^0}{\ell}}{q^2\sqrt{2} k \cdot p_{+} \sqbr{\ell}{  \tilde{p}_{-}} \agbr{ \ell }{\tilde{p}_{+}} },\notag\\
	\alpha^-_{22}  = -(\alpha^+_{11})^* &  \simeq 0 ~,\label{eqn:BHSALO}
\end{align}
in which we have dropped the photon subscripts, and uniform overall signs or factors of $i$;  $k,\ell$ = $k_1, k_2$ or $k_2,k_1$ for photon 1 and 2 respectively; and the spinor notation is detailed in Appendix \ref{app:SHF}. 
The parity relations \eqref{eqn:PA} are satisfied as expected. Eqs.~\eqref{eqn:HBHA} and \eqref{eqn:BHSALO} together provide a compact expression of the leading order HBH square amplitude. 

It should be understood that the particular form for the spin helicity amplitudes above depends on the choice of reference momenta, because the amplitudes explicitly depend on polarization vectors and spinors (see App. \ref{app:SHF}).  Moreover, the ability to straightforwardly expand the full results to the leading order results depends on a sensible choice of reference momenta. In contrast, the full unpolarized BH rate is independent of polarizations and spinors, as a result of spinor and polarization vector completeness, and therefore must be independent of any such reference momenta choice. 

\section{Polarization-decomposed HBH rate}
Here we give the explicit results of a Higgs-Bethe-Heitler Feynman type calculation. To preempt the loss of numerical precision from large cancellations due to the Ward identity, we do not use polarization completeness relations. Rather, we retain the polarization vectors explicitly in the HBH rate. In the case that the Higgs is at rest in the lab frame, we simply use a Cartesian basis for the polarization vectors, aligning the back-to-back photons with the $z$-axis. That is, in the HBH square amplitude we coherently sum over the polarization basis
\begin{equation}
	\label{eqn:PDHBHPD}
	\epsilon^1_\mu(k_{1,2}) = (0,1,0,0)\quad \mbox{and} \quad \epsilon^2_\mu(k_{1,2}) = (0,0,1,0)~.
\end{equation}

The squared matrix element for the process $h+N_1+N_2\rightarrow\gamma(k_1) \gamma(k_2)+N_1+N_2\rightarrow 4\ell+N_1^\prime+N_2^\prime$ is given by
\begin{equation}
|\Em|^2 = \mcA^{\mu\nu}\mcA^{*\alpha\beta} \!\!\!
\mathop{\sum_{\text{pols}}}_{ \text{a,b,c,d}}\Big[\epsilon^{a*}_\mu(k_1)\epsilon^{b*}_\nu(k_2) \epsilon^c_\alpha(k_1)\epsilon^d_\beta(k_2) \epsilon^a_{\mu^\prime}(k_1)\epsilon^{c*}_{\alpha^\prime}(k_1) \epsilon^b_{\nu^\prime}(k_2)\epsilon^{d*}_{\beta^\prime}(k_2)\Big]
\BH_1^{\mu^\prime\alpha^\prime}\,\BH_2^{\nu^\prime\beta^\prime}
\end{equation}
 where the tensor $\mcA^{\mu\nu}$ is 
 \begin{equation}
\mcA^{\mu\nu}=
  c\left(k_1\cdot k_2\,g^{\mu\nu}-k_2^\mu k_1^\nu\right)+
  \tilde c\,\epsilon^{\alpha\mu\beta\nu}k_{1\alpha} k_{2\beta}~.
\end{equation}
The $\BH_i$ factors are the polarized Bethe-Heitler squared amplitudes for a photon $i$, including form factor contributions, these are in general
\begin{align}
 \BH^{ab}
 &=
 \frac{e^6\,\mathcal{G}}{q^4(k\cdot p_-)^2 (k\cdot p_+)^2} \Bigg\{
  \Big[2(k\cdot p_-)(k\cdot p_+)\left(2E_- E_+-m^2-p_+\cdot p_-\right)\mathcal{B}^{ab}_{k,p_-}\notag\\
  &+\frac{8E_+}{M}(k\cdot p_+)^2(k\cdot p_-)\mathcal{B}^{ab}_{P,p_-}+(k\cdot p_+)^2(4E_+^2+q^2)\mathcal{F}^{ab}_{p_-} + (p_+ \leftrightharpoons p_-)\Big]\notag\\
  &-8(k\cdot p_-)^2(k\cdot p_+)^2\frac{P^aP^b}{M^2}\notag\\
  &-\frac{2}{M}(k\cdot p_-)(k\cdot p_+)\left[-q^2+2\left(E_-(k\cdot p_-)+E_+(k\cdot p_+)\right)\right]\mathcal{B}^{ab}_{P,k}\notag\\
  &+2(k\cdot p_-)(k\cdot p_+)(q^2-4E_-E_+)\mathcal{B}^{ab}_{p_+,p_-}\notag\\
  &+2(k\cdot p_-)(k\cdot p_+) \left[\left(k\cdot p_-+k\cdot p_+\right)^2+q^2E_\gamma^2\right]g^{ab}\notag\\
  &-4(k\cdot p_-)(k\cdot p_+) \left(2E_-E_+-m^2-p\cdot r\right)k^ak^b\Bigg\}~,
\label{eqn:PolDec}
\end{align}
where $\mathcal{F}^{ab}_{\ell} = \ell^a k^b + \ell^b k^a - 2\ell^a \ell^b$ and $\mathcal{B}^{ab}_{\ell_1\ell_2} = \ell_1^a \ell^b_2 + \ell_1^b \ell^a_2$. The photon momentum is denoted by $k$ and the lepton momenta by $p_+,\,p_-$. $M$ and $m$ are the masses of the nucleus and lepton respectively. Assuming that the Higgs is at rest in the lab frame and that the photon is in the $z$-direction, as in Eq.~\eqref{eqn:PDHBHPD}, the previous expression simplifies to
\begin{equation}
  \BH^{ab}\simeq \frac{2\,e^{6}\,\mathcal{G}}{q^4}\left\{
		\frac{g^{ab}\left[E_\gamma^2\,q^2+(k\cdot p_- + k\cdot p_+)^2\right]}{(k\cdot p_-)(k\cdot p_+)}
		-4\left(\frac{E_{p_+}p_-^a}{k\cdot p_-}+\frac{E_{p_-}p_+^a}{k\cdot p_+}\right)
			\left(\frac{E_{p_+}p_-^b}{k\cdot p_-}+\frac{E_{p_-}p_+^b}{k\cdot p_+}\right)
	\right\}~,
\label{eqn:PolDecS}
\end{equation}
\noindent where, in the second term, we expanded terms of the form $\sqrt{4E^2\pm q^2}$ to leading order in $q^2/E^2$. We have checked that the helicity formalism results and the Feynman diagram calculation results for HBH rate agree. 

Finally, the unpolarized BH rate can be obtained from Eq.~\eqref{eqn:PolDecS} simply by averaging over the photon polarization as follows
\begin{equation}
	\langle \left|\Em\right|^2\rangle = 1/2\left(\BH^{11}+\BH^{22}\right)~.
\end{equation}
It is instructive to use the polarization vector completeness relation to obtain an expression in terms of Lorentz dot products. In this case, starting with Eq.~\eqref{eqn:PolDec}, the BH rate is given by
\begin{align}
  \langle \left|\Em\right|^2\rangle=&2\,e^{6}\,\mathcal{G}\Bigg(-2\frac{\left(E_-^2+E_+^2\right)}{q^2\,k\cdot p_-\,k\cdot p_+} 
	+4\frac{m^2 \left(k\cdot p_- E_--k\cdot p_+ E_+\right)^2}{q^4\,(k\cdot p_-)^2\,(k\cdot p_+)^2}\nonumber\\
	&+\frac{m^2 \left(k\cdot p_-+k\cdot p_+\right)^2}{q^2\,(k\cdot p_-)^2\,(k\cdot p_+)^2}
	- \frac{k\cdot p_- k\cdot p_+}{(k\cdot p_-)^2\,(k\cdot p_+)^2}
	-2 \frac{\left(k\cdot p_-\right){}^2+\left(k\cdot p_+\right)^2}{q^4\,(k\cdot p_-)\,(k\cdot p_+)}\nonumber\\
  &-2\frac{k\cdot p_-+k\cdot p_+}{q^2\,k\cdot p_-\,k\cdot p_+}\Bigg)~,
\end{align}
\noindent where the leading terms -- the terms on the first line -- reproduce Eq.~\eqref{eqn:BHLOR}.

\section{Numerical simulations of Bethe-Heitler conversion}
\label{app:BHDR}
In this appendix we present numerical evaluations of several differential BH rates. The numerics were done in two ways: by numerically integrating the full tree-level analytical results -- i.e. the BH rate arising from Eqs.~\eqref{eqn:BHSA} or \eqref{eqn:PolDec} with appropriate integration measures -- using the CUBA library \cite{Hahn:2004fe}; and with a Monte Carlo (MC) code developed privately (the details are given in Appendix \ref{app:MCgen}). Fig. \ref{fig:enspec} shows the differential distribution of the positron energy fraction $\mathcal{E}_+=E_+/E_\gamma$. For efficiency the MC  simulation (blue binned histogram) is generated with a cut on the difference of electron and positron azimuthal angles, $\delta\phi\equiv(\phi_{+}-\phi_{-}) \mod 2\pi \in [0.6\pi,1.4\pi]$ (see App. \ref{app:MCgen}). This agrees with the numerical integration (red line) for the same cuts on $\delta\phi$. 

Fig.  \ref{fig:enspec} (right) shows the positron energy distribution after applying the opening angle cut of $\theta_{\ell\ell}>10^{-4}$ on the angle between $e^+$ and $e^-$ momenta. The MC agrees with the full numerical integration of the BH rate even though the $\delta\phi$ cut is still applied in the generation of the events
. Fig.~\ref{fig:enspec} demonstrates that the asymmetric configurations, where one of the two leptons carries the larger part of the photon energies, are the more probable ones, especially for non-zero opening angles. 

\begin{figure}
\centering
\includegraphics[width=0.49\columnwidth]{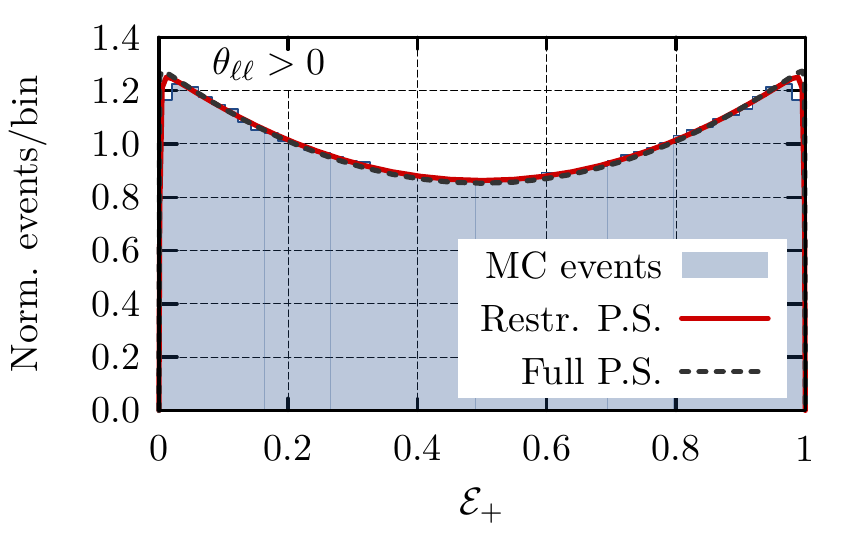}
\centering
\includegraphics[width=0.49\columnwidth]{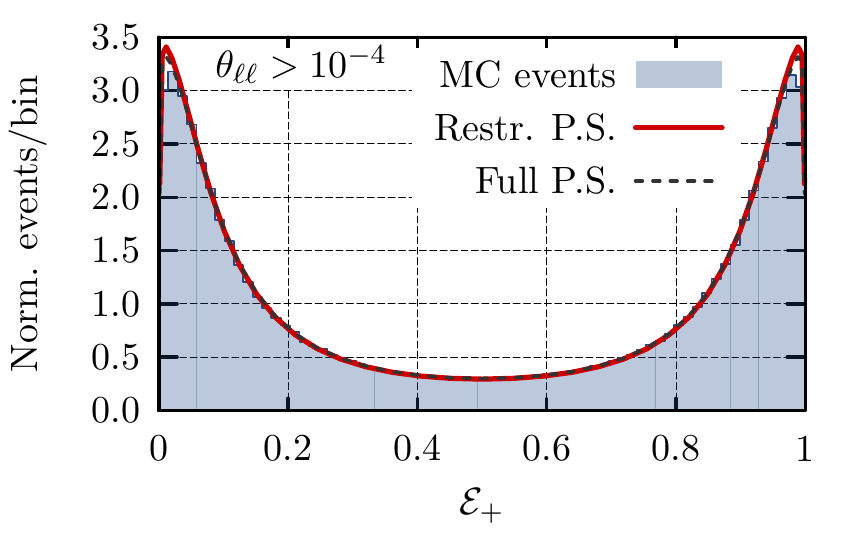}
	\caption{Spectrum of the positron energy $\mathcal{E}_+ = E_+/E_\gamma$. No opening angle cut was applied in the left hand figure and an opening angle cut of $10^{-4}$ was applied in the right hand one. The histograms were created with MC events and the solid curves are results of numerically integrating the differential cross section. The dashed curve in both figures is the result of numerically integrating the differential cross section over the entire range of $\delta\phi$ as opposed over the range $[0.6\pi,1.4\pi]$. }
	\label{fig:enspec}
\end{figure}

\begin{figure}
\centering
\includegraphics[width=0.49\columnwidth]{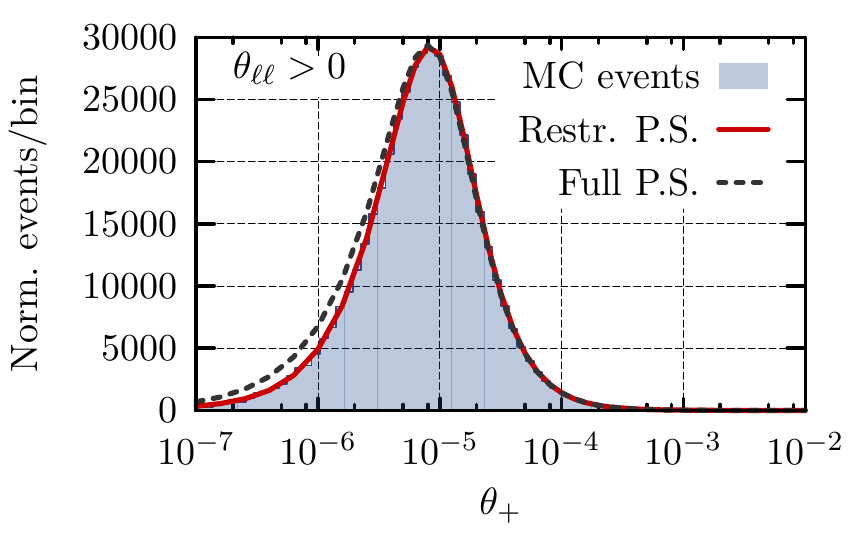}
\centering
\includegraphics[width=0.49\columnwidth]{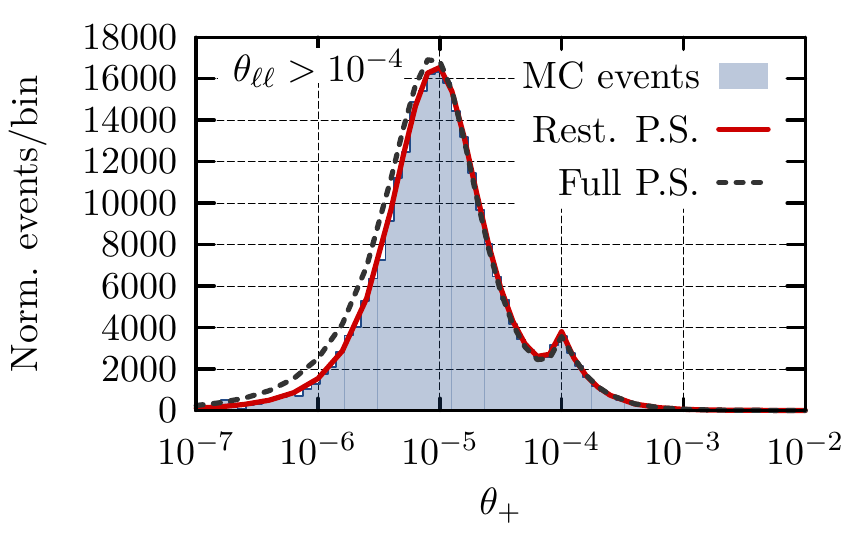}
	\caption{Polar angle distribution of the leptons. No opening angle cut was applied in the left hand figure and an opening angle cut of $10^{-4}$ was applied in the right hand one. The histograms were created with MC 	events and the solid curves are results of numerically integrating the differential rate expression. The small bump in the right hand figure $(\sim 10^{-4})$ is a result of applying an opening angle cut. Its location is a function of the cut.
	}
	\label{fig:poldist}
\end{figure}

Fig. \ref{fig:poldist} shows the positron polar angle distribution and demonstrates the combined effect of the Si nuclear form factor and the smaller available phase space that suppress very small momentum transfers and thus very small polar angles. The peak is at $\sim m/E\sim 10^{-5}$ both for the distribution without a cut on the $e^+e^-$ opening angle $\theta_{\ell\ell}$, Fig. \ref{fig:poldist} (left), and for the case where $\theta_{\ell\ell}>10^{-4}$ is imposed, Fig. \ref{fig:poldist} (right). This cut also results in an additional peak in the distribution, cf. Fig. \ref{fig:poldist} (right).

\begin{figure}
	\centering \includegraphics[width=0.49\columnwidth]{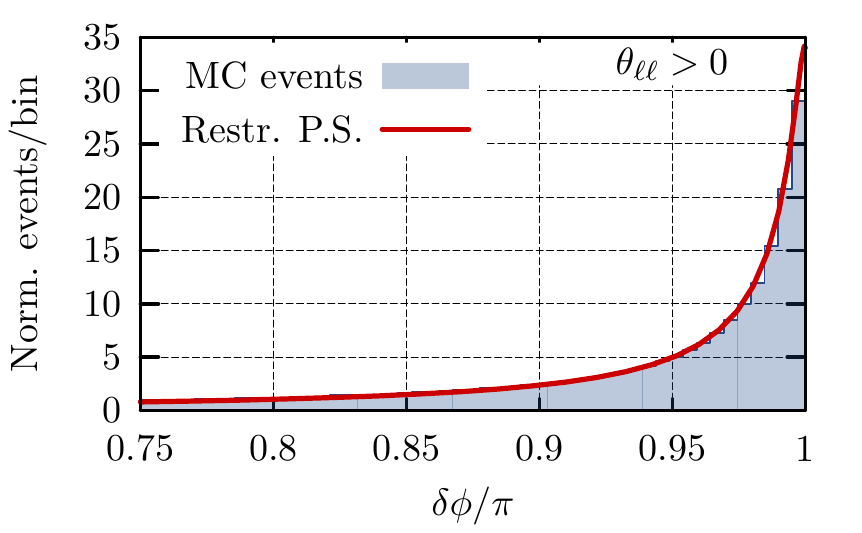}
	\centering \includegraphics[width=0.49\columnwidth]{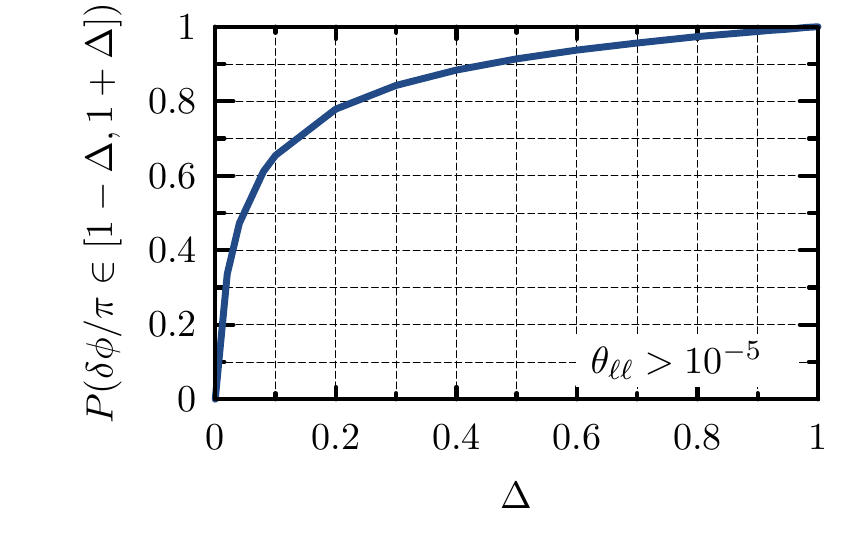}
	\caption{Left: Distribution of the azimuthal angle between the leptons. The histograms were created with MC events and the solid curves are results of numerically integrating the differential rate expression. Right: The cumulative distribution function of the relative azimuthal orientation, $P(\delta\phi/\pi \in [1- \Delta, 1+ \Delta])$, from numerical integration. 
	}
	\label{fig:azimdist}
\end{figure}

In Fig. \ref{fig:azimdist}, the distribution in the relative azimuthal angle $\delta\phi$ of the two leptons is shown. The majority of the events are close to the coplanar configuration, where the photon and the two lepton momenta all lie in the same plane. However, it is noteworthy that approximately $40\%$ of BH events have acoplanarity of $\sim5\%$ or more.

\section{Analysis for \texorpdfstring{$q\bar q\to \gamma\gamma$}{qqgg}}

Performing the measurement proposed in this work faces two main challenges: first, resolving and reconstructing the electron and positron directions; and second identifying a background-poor sample of events with Higgs decaying to diphotons. Regarding the first challenge, one might simply ask how well and with what efficiency can the LHC or a future collider detector reconstruct the details of photon conversion. To do so, the experimental collaborations may wish to test the polarization structure of a standard model (non-Higgs) amplitude. To demonstrate that there is a non-trivial structure to be measured in SM conversions, we briefly analyze here the leading production of diphotons at the LHC.

The dominant  diphoton production (and dominant background for Higgs to photons events) is $\qqgg$ scattering. This has tree-level spin-helicity amplitude
\begin{equation}
	[\mathcal{M}_{\rm BG}]^{\lambda_1\lambda_2}_{rs} = 
\parbox[c]{6 cm}{
\includegraphics{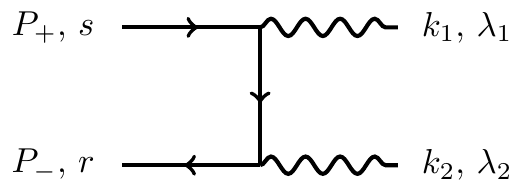}
} + ~\quad \mbox{u channel}~,
\end{equation}
in which $r = 1,2$ ($s = 1,2$) is  the spin of the (anti)-fermion, and $\lambda_{1,2} = \pm$ are the usual photon helicities. The two photons then convert in the tracker.
The background rate can be written in a form similar to Eq.~\eqref{eqn:HBHA}. That is, for one fermion species
\begin{equation}
	|\mathcal{M}_{\rm BG}|^2 = \mcG_1\mcG_2\sum_{r,s} \sum_{r_1,s_1,r_2,s_2} \Big| \sum_{\lambda_1,\lambda_2} [\mathcal{M}_{\rm BG}]^{\lambda_1\lambda_2}_{rs}\alpha^{\lambda_1}_{1_\rs} \alpha^{\lambda_2}_{2_\rs}\Big|^2~.
\end{equation}	
As is well-known, the only non-zero independent amplitudes are $[\mathcal{M}_{\rm BG}]^{+ -}_{12}$ and $[\mathcal{M}_{\rm BG}]^{-+}_{12}$. 
One finds with our usual choice of reference momenta, and the light cone decomposition \eqref{eqn:SLCD} for the quark momenta $P_\pm$,
\beq
\label{eqn:BGC}
\begin{split}
	[\mathcal{M}_{\rm BG}]^{+ -}_{12} = \big([\mathcal{M}_{\rm BG}]^{ -+}_{21}\big)^* & = \frac{\mathcal{A}_+(k_2,k_1)}{P_+\cdot k_1} - \frac{\mathcal{A}_-^*(k_1,k_2)}{P_-\cdot k_1}~,\\
	[\mathcal{M}_{\rm BG}]^{-+}_{12} = \big([\mathcal{M}_{\rm BG}]^{ +-}_{21}\big)^* & = \frac{\mathcal{A}_+(k_1,k_2)}{P_+\cdot k_1} - \frac{\mathcal{A}_-^*(k_2,k_1)}{P_-\cdot k_1}~,\\
	\mathcal{A}_{\pm}(k,\ell) & = Q_f^2\frac{\agbr{\tilde{P}_\pm}{\ell}^2 \sqbr{\tilde{P}_\pm}{k} \sqbr{\tilde{P}_\mp}{k}}{2 k_1 \cdot k_2} ~,
	\end{split}
\eeq
with $Q_f$ the fermion electric charge. The square amplitude simplifies to
\begin{equation}
	\label{eqn:BGR}
	|\mathcal{M}_{\rm BG}|^2 = 2\mcG_1\mcG_2\sum_{r_1,s_1,r_2,s_2} \Big|[\mathcal{M}_{\rm BG}]^{+-}_{12}\alpha^+_{1_\rs}\alpha^-_{2_\rs} + [\mathcal{M}_{\rm BG}]^{-+}_{12}\alpha^-_{1_\rs}\alpha^+_{2_\rs}\Big|^2~.
\end{equation}
It is interesting to contrast this with the HBH result. Here the helicity interference arises in terms of the form $\alpha^-_{1_{12}}\alpha^+_{2_{12}}\alpha^-_{1_{21}}\alpha^+_{2_{21}}$, rather than from $\alpha^-_{1_{12}}\alpha^-_{2_{12}}\alpha^-_{1_{21}}\alpha^-_{2_{21}}$ as we found for HBH. Since $\alpha^+_{12} = (\alpha^-_{21})^*$, etc, it follows from the explicit results \eqref{eqn:BHSALOA} that this helicity flip on branch `2'  produces a phase change $\phi_{2_\pm} \to -\phi_{2_\pm}$ in the background interference term, compared to the HBH interference terms. 

To compare with the Higgs rest frame HBH rate \eqref{eqn:GVPD}, we assume the quark centre of mass frame aligns with the nuclear rest frame. Integrating all over azimuthal structure except $\psi \equiv \phi_{1_+} - \phi_{2_+}$ (cf. $\varphi = \phi_{1_+} + \phi_{2_+}$ in Eq.~\eqref{eqn:DVP}), one finds that the background rate has the form
\begin{equation}
	\frac{d\Gamma}{d\psi \dPS_{\gamma,\theta}} = \mathcal{A}^{\rm BG}_{\gamma,\theta} + \mathcal{B}^{\rm BG}_{\gamma,\theta} \cos( 2\psi)~.
\end{equation}
We see that this differential rate has sinusoidal dependence on the mean azimuthal orientation, $\psi$, of the outgoing positrons, with respect to, say, the incoming quarks -- the beamline --  rather than the inter-branch lepton azimuthal orientation, $\varphi$, as in HBH. Moreover, the phase change $\phi_{2_\pm} \to -\phi_{2_\pm}$ ensures background is flat in $\varphi$. Note also that unlike the HBH process, the $[\mathcal{M}_{\rm BG}]^{+-}$ factors ensure this background rate features higher spin waves, so that its angular differential structure will differ from the HBH structure, too.

In summary, the leading-order doubly-converted $q\bar{q} \to \gamma\gamma$ square amplitude is given explicitly by Eqs.~\eqref{eqn:BGC} and \eqref{eqn:BGR} combined with the BH spin-helicity amplitudes \eqref{eqn:BHSALO}. In the $q\bar{q}$ center of mass frame, the corresponding leading order BH amplitudes are given in Eqs.~\eqref{eqn:BHSALOA}.

\section{Monte Carlo numerical schemes}
\label{app:MCgen}
To generate Monte Carlo (MC) events for the HBH process, we first generate unpolarized BH events and then use the von Neumann rejection technique to re-weight the events according to the HBH differential rate. For a single HBH event one needs two BH events taken from disjoint MC samples. Therefore, we first describe the generation of unpolarized BH events.

The phase space for a single BH event, $\gamma N\to e^+e^- N$, is five dimensional. We take $z$-axis to be the incoming photon direction. For conversion of unpolarized photons, the kinematics are invariant under overall azimuthal rotations around the $z$ axis. We therefore fix the positron azimuthal angle to zero. The remaining four coordinates are chosen to be the electron energy fraction $\mathcal{E}_-= E_-/E_\gamma$, two transformed polar angles, $t_1=\log_{10}(\theta_+)$ and $t_2=\log_{10}(\theta_-)$, and the azimuthal angle of the electron $\phi_-$ (see Fig. \ref{fig:PAD} for definitions). 

For MC we used two independent private codes. One is written in \verb|C| and the other in \texttt{C++/Java}. To populate the BH phase space we first randomly generate the values for $\mathcal{E}_-$, $t_{1,2}$ and $\phi_-$ according to either uniform distributions or conveniently chosen initial probability density functions (PDFs) and then unweight to obtain the BH event distribution. 

For the \verb|C| code, the initial PDFs are as follows. For a $60$ GeV photon, generating $\mathcal{E}_-$ according to a uniform distribution results in an efficiency of $\sim70\%$. We therefore generate a uniform distribution of $\mathcal{E}_- \in [m/E_\gamma,1-m/E_\gamma]$. The transformed polar angle variables, $t_1$ and $t_2$, are generated according to uniform distributions in a suitable numerical range, see Table \ref{Table:MC}. Using the coordinates $t_{1,2}$ captures the fact that electron and positron distributions are sharply peaked around $\theta_\pm \sim m/E$. The BH events are also dominated by kinematic configurations that are not too far from the coplanar one. We therefore generate $\phi_-$ in the range $[0.6\pi,1.4\pi]$, which suffices for our precision. To capture the fact that the BH distribution is peaked toward $\phi_-/\pi=1$, the \verb|C| code generates $\phi_-$ according to a Cauchy distribution (Lorentzian) with location parameter $x_o=\pi$ and scale parameter $\lambda=0.03\pi$~\cite{johnson1994distributions} to improve the efficiency. The \texttt{C++/Java} uses similar initial data, with the exception of slightly different ranges and that $\phi_-$ is populated by a uniform distribution on a slightly narrowed domain $\phi_- \in [3\pi/4, 5\pi/4]$ (see Table. \ref{Table:MC}).

\begin{table}[t]\centering
\begin{tabular}{c|cc|cc}
\hline\hline 
 & \multicolumn{2}{c|}{\tt{C}} & \multicolumn{2}{c}{\texttt{C++/Java}} \\
Parameter & Range & PDF & Range & PDF\\ 
\hline
$\mathcal{E}_-$ & $[m/E_\gamma,1-m/E_\gamma]$ & Uniform & $[m/E_\gamma,1-m/E_\gamma]$ & Uniform\\ 
$t_1$ & $[-7,-2]$ & Uniform & $[-6,-3]$ & Uniform \\ 
$t_2$ & $[-7,-2]$ & Uniform & $[-6,-3]$ & Uniform \\ 
$\phi_-$ & $[0.6\pi,1.4\pi]$ & Lorentzian & $[3\pi/4,5\pi/4]$ & Uniform \\ 
\hline\hline 
\end{tabular} 
\caption{The details on the MC generation of BH events, with phase space variables (1st column) for \texttt{C} (\texttt{C++/Java}) generated in the range given in the 2nd (4th) column according to the distribution given in the 3rd (5th) column (for details see text).}
 \label{Table:MC}
\end{table}

In the next step we unweight the events generated from initial PDFs to obtain the proper BH distribution. In the unweighting, the events are rejected with a probability that is $1-w$, with $w=(d\Gamma_{\text{\sc bh}}/d\text{\sc ps})/\max(d\Gamma_{\text{\sc bh}}/d\text{\sc ps})$.
The BH MC event sample was validated by comparing the generated event distributions to the results of numerical integration of BH differential cross-sections as shown in Figs.~\ref{fig:enspec},~\ref{fig:poldist}~and~\ref{fig:azimdist}. Additionally, in Figs.~\ref{fig:enspec}~and~\ref{fig:poldist}, the distributions with a cut on the opening angle between $e^+$ and $e^-$ of $\theta_{\ell\ell}>10^{-4}$ are shown. The MC sample is in excellent agreement with the results of numerical integration.

In the final step, we convert the generated BH events into MC event samples for the HBH process. To do so, two disjoint BH samples were used -- one sample per photon branch. The rate for two BH events is given by $(d\Gamma_{\text{\sc bh}_1}/d\text{\sc ps})(d\Gamma_{\text{\sc bh}_2}/d\text{\sc ps})$. To obtain the proper HBH even rates, we use the standard reweighting technique where events are rejected according to the weight
\begin{equation}
	w=\frac{(d\Gamma_\text{\sc hbh}/d\text{\sc ps})[\varphi]}{(d\Gamma_{\text{\sc bh}_1}/d\text{\sc ps})\,(d\Gamma_{\text{\sc bh}_2}/d\text{\sc ps})}~,
\end{equation}
where the twist angle between the positrons, $\varphi$, is populated by a uniform distribution on $[0,2\pi]$. 

The two MC codes have been cross tested. In numerics we use $3\times 10^6$ HBH events from the \verb|C| generator and $8 \times 10^5$ HBH events from the \texttt{C++/Java} MC generator.


\providecommand{\href}[2]{#2}\begingroup\raggedright\endgroup

\end{document}